\documentclass[twoside]{ae100prg}
\usepackage{graphicx}
\usepackage[breaklinks]{hyperref}
\usepackage{booktabs}

\newcommand{\vk}{v_{\rm k}}
\newcommand{\vb}{v_\mathrm{b}/c}
\newcommand{\ms}{{\rm ms}}
\newcommand{\ks}{\,{\rm km/s}}

\newcommand{\MeV}{\,{\rm MeV}}

\newcommand{\cf}{\textit{cf.,}~}
\newcommand{\ie}{\textit{i.e.,}~}
\newcommand{\eg}{\textit{e.g.,}~}
\newcommand{\newcapt}[1]{\caption{\scriptsize{#1}}}
\newcommand{\newfoot}[1]{\footnote{\scriptsize{#1}}}

\font\tenscr=rsfs10 scaled1100
\font\sevenscr=rsfs7 
\font\fivescr=rsfs5 
\skewchar\tenscr='177
\skewchar\sevenscr='177
\skewchar\fivescr='177
\newfam\scrfam
\textfont\scrfam=\tenscr
\scriptfont\scrfam=\sevenscr
\scriptscriptfont\scrfam=\fivescr

\def\scri{{\fam\scrfam I}}

\begin{document}
\title[Three little pieces]{Three little pieces for computer and relativity}

\author{Luciano Rezzolla}

\address{Max-Planck-Institut f\"ur Gravitationsphysik, Albert-Einstein-Institut, Potsdam, Germany}

\begin{abstract}
Numerical relativity has made big strides over the last decade. A number
of problems that have plagued the field for years have now been mostly
solved. This progress has transformed numerical relativity into a
powerful tool to explore fundamental problems in physics and
astrophysics, and I present here three representative examples. These
``three little pieces'' reflect a personal choice and describe work that
I am particularly familiar with. However, many more examples could be
made.
\end{abstract}

\section{Introduction}

Numerical relativity has hardly seen better times before. Over the last
few years, in fact, a truly remarkable development has shaken the
field. Starting from the first simulations showing that black-hole
binaries could be evolved for a few orbits~\cite{Pretorius:2005gq,
  Campanelli:2005dd, Baker:2005vv}, or that black-hole formation could be
followed stably using simple gauges and without
excision~\cite{Baiotti06}, new results, some awaited for decades, have
been obtained steadily. As a direct consequence of this ``Renaissance'',
it is now possible to simulate binary black holes~\cite{Chu:2009md} and
binary neutron stars~\cite{Baiotti08} accurately for dozens of orbits,
from the weak-field inspiral, down to the final black-hole ringdown (see
also~\cite{Centrella:2010,Pfeiffer2012} for recent reviews).

There are several reasons behind this rapid progress. These include the
use of more advanced and accurate numerical
techniques~\cite{Radice2011,Radice2012a}, the availability of larger
computational facilities, but also the development of formulations of the
Einstein equations and gauges that are particularly well-suited for
numerical evolutions~\cite{Nakamura87, Bona94b, Shibata95, Baumgarte99,
  Alcubierre02a, Pretorius:2004jg, Alic:2009, Bernuzzi:2009ex,
  Mueller:2010bu, Schnetter:2010cz, Alic:2011a}. The phase transition
that has taken during this year has radically changed numerical
relativity, freeing it from the corner of idealised investigations. Most
importantly, it has transformed numerical relativity into a research area
where long-standing problems can found a quantitative and accurate
solution, and into a tool by means of which it is possible to
\textit{explore} fundamental aspects of physics and astrophysics.

Numerous examples could be given to testify this transformation, although
I will report here only those that I am particular familiar with. More
specifically, in what follows I will discuss: \textit{(i)} how numerical
simulations of magnetised neutron stars provide convincing evidence that
this process leads to the conditions that are expected behind the
phenomenology of short gamma ray burst; \textit{(ii)} how numerical
simulations of the head-on collision of selfgravitating fluids boosted at
relativistic speeds can be used to understand the conditions leading to
the formation of a black hole and provide a dynamical version of the hoop
conjecture; \textit{(iii)} how the study of the local properties of
apparent horizons can be used to explain bizarre behaviours in binary
black-hole simulations and can be effectively correlated with a portion
of the spacetime infinitely far away: $\scri^+$. This selection is by no
means comprehensive, but rather a very personal one, and I apologise in
advance for not discussing all the excellent work that cannot find space
in this contribution.

\section{First piece: From neutrons star to gamma-ray bursts}

The numerical investigation of the inspiral and merger of binary neutron
stars in full general relativity has seen enormous progress made in recent
years. Crucial improvements in the formulation of the equations and
numerical methods, along with increased computational resources, have
extended the scope of early simulations. These developments have made it
possible to compute the full evolution, from large binary-separations up
to black-hole formation, without and with magnetic
fields~\cite{Shibata06a, Baiotti08, Anderson2008, Liu:2008xy,
  Giacomazzo:2009mp, Giacomazzo2011}, and with idealised or realistic
equations-of-state \cite{Rezzolla:2010,Kiuchi2010}. This tremendous
advancement is also providing information about the entire gravitational
waveform, from the early inspiral up to the ringing of the black hole
(see, \eg~\cite{Baiotti:2010, Bernuzzi2012, Hotokezaka2013b}). Advanced
interferometric detectors starting from 2014 are expected to observe
these sources at a rate of $\sim40-400$ events per
year~\cite{Abadie:2010}.

These simulations also probe whether the end-product of mergers can serve
as the ``central engine'' of short gamma-ray bursts
(SGRBs)~\cite{Paczynski86,Eichler89,Narayan92}. The prevalent scenario
invoked to explain SGRBs involves the coalescence of a binary system of
compact objects, \eg a black hole and a neutron star or two neutron
stars~\cite{Ruffert99b,Rosswog:2003,Nakar:2007yr,Lee:2007js}. After the
coalescence, the merged object is expected to collapse to a black hole
surrounded by an accretion torus. An essential ingredient in this
scenario is the formation of a central engine, which is required to
launch a relativistic outflow with an energy of $\sim10^{48}-10^{50}$~erg
on a timescale of $\sim0.1-1$~s~\cite{Nakar:2007yr,Lee:2007js}.

The qualitative scenario described above is generally supported by the
association of SGRBs with old stellar populations, distinct from the
young massive star associations for long
GRBs~\cite{Fox2005,Prochaska2006}. It is also supported to a good extent
by fully general-relativistic simulations, which show that the formation
of a torus of mass $M_{\rm{tor}}\lesssim0.4\,M_{\odot}$ around a black
hole with spin $J/M^2\simeq0.7-0.8$, is
inevitable~\cite{Rezzolla:2010}. In addition, recent simulations have
also provided the first evidence that the merger of a binary of modestly
magnetised neutron stars naturally forms many of the conditions needed to
produce a jet of ultrastrong magnetic field, with properties that are
broadly consistent with SGRB observations. This \textit{missing link}
between the astrophysical phenomenology of GRBs and the theoretical
expectations is a genuine example of the new potential of numerical
relativity and I will discuss it in detail below\newfoot{Much of what
  follows is taken from the discussion presented in
  Ref.~\cite{Rezzolla:2011}.}.

\subsection{The numerical setup}

It is not useful to discuss here in detail the numerical setup and the
technical details of the numerical codes used in these
calculations. These details can be found in
Refs.~\cite{Giacomazzo:2007ti,Giacomazzo2011}, while a description of the
physical initial data was presented in Ref.~\cite{Rezzolla:2011}. It is
sufficient to recall here that the evolution of the spacetime is obtained
using a three-dimensional finite-differencing code providing the solution
of a conformal traceless formulation of the Einstein
equations~\cite{Pollney:2007ss} (\ie the \texttt{CCATIE} code). The
equations of general-relativistic magnetohydrodynamics (GRMHD) in the
ideal-MHD limit are instead solved using a code
code~\cite{Baiotti03a,Baiotti04,Giacomazzo:2007ti} which adopts a
flux-conservative formulation of the equations as presented
in~\cite{Anton05} and high-resolution shock-capturing schemes (\ie the
\texttt{Whisky} code). In order to guarantee the divergence-free
character of the MHD equations the flux-CD approach described
in~\cite{Toth2000} was employed, although with the difference that the
vector potential is used as evolution variable rather than the magnetic
field. Both the Einstein and the GRMHD equations are solved using the
vertex-centred adaptive mesh-refinement (AMR) approach provided by the
\texttt{Carpet} driver~\cite{Schnetter-etal-03b}. In essence, the
highest-resolution refinement level is centred around the peak in the
rest-mass density of each star and in moving the ``boxes'' following the
position of this maximum as the stars orbit. The boxes are evolved as a
single refinement level when they overlap. The calculations were carried
out using six levels of mesh refinement with the finest level having a
resolution of $\Delta = 0.1500\,M_{\odot}\simeq 221\,\mathrm{m}$.

From a physical point of view, the initial data is given by a
configuration that could represent the properties of a neutron
star-binary a few orbits before their coalescence. More specifically, we
simulate two equal-mass neutron stars, each with a gravitational mass of
$1.5\,M_{\odot}$ (\ie sufficiently large to produce a black hole soon
after the merger), an equatorial radius of $13.6\,{\rm km}$, and on a
circular orbit with initial separation of $\simeq45\,{\rm km}$ between
the centres (all lengthscales are coordinate scales). Confined in each
star is a poloidal magnetic field with a maximum strength of
$10^{12}\,{\rm G}$. At this separation, the binary loses energy and
angular momentum via emission of gravitational waves, thus rapidly
proceeding on tighter orbits as it evolves.

\subsection{The basic dynamics}

After about $8\,\ms$ ($\sim3$ orbits) the two neutron stars merge forming
a hypermassive neutron star (HMNS), namely, a rapidly and
differentially-rotating neutron star, whose mass, $3.0\,M_{\odot}$, is
above the maximum mass, $2.1\,M_{\odot}$, allowed with uniform rotation
by an ideal-fluid equation of state (EOS)\newfoot{The use of a
  simplified EOS does not influence particularly the results besides
  determining the precise time when the HMNS collapses to a black hole.},
$p=(\Gamma-1)\rho\epsilon$, where $\rho$ is the baryonic density,
$\epsilon$ the specific internal energy, and $\Gamma=2$ with an adiabatic
index of 2. Being metastable, a HMNS can exist as long as it is able to
resist against collapse via a suitable redistribution of angular momentum
[\eg deforming into a ``bar'' shape \cite{Shibata06a,Baiotti08}], or
through the increased pressure-support coming from the large
temperature-increase produced by the merger. However, because the HMNS is
also losing angular momentum through gravitational waves, its lifetime is
limited to a few ms, after which it collapses to a black hole with mass
$M=2.91\,M_{\odot}$ and spin $J/M^2=0.81$, surrounded by a hot and dense
torus with mass $M_{\rm{tor}}=0.063\,M_{\odot}$~\cite{Giacomazzo2011}.

\begin{figure*}
\begin{center}
     \includegraphics[angle=0,width=5.5cm]{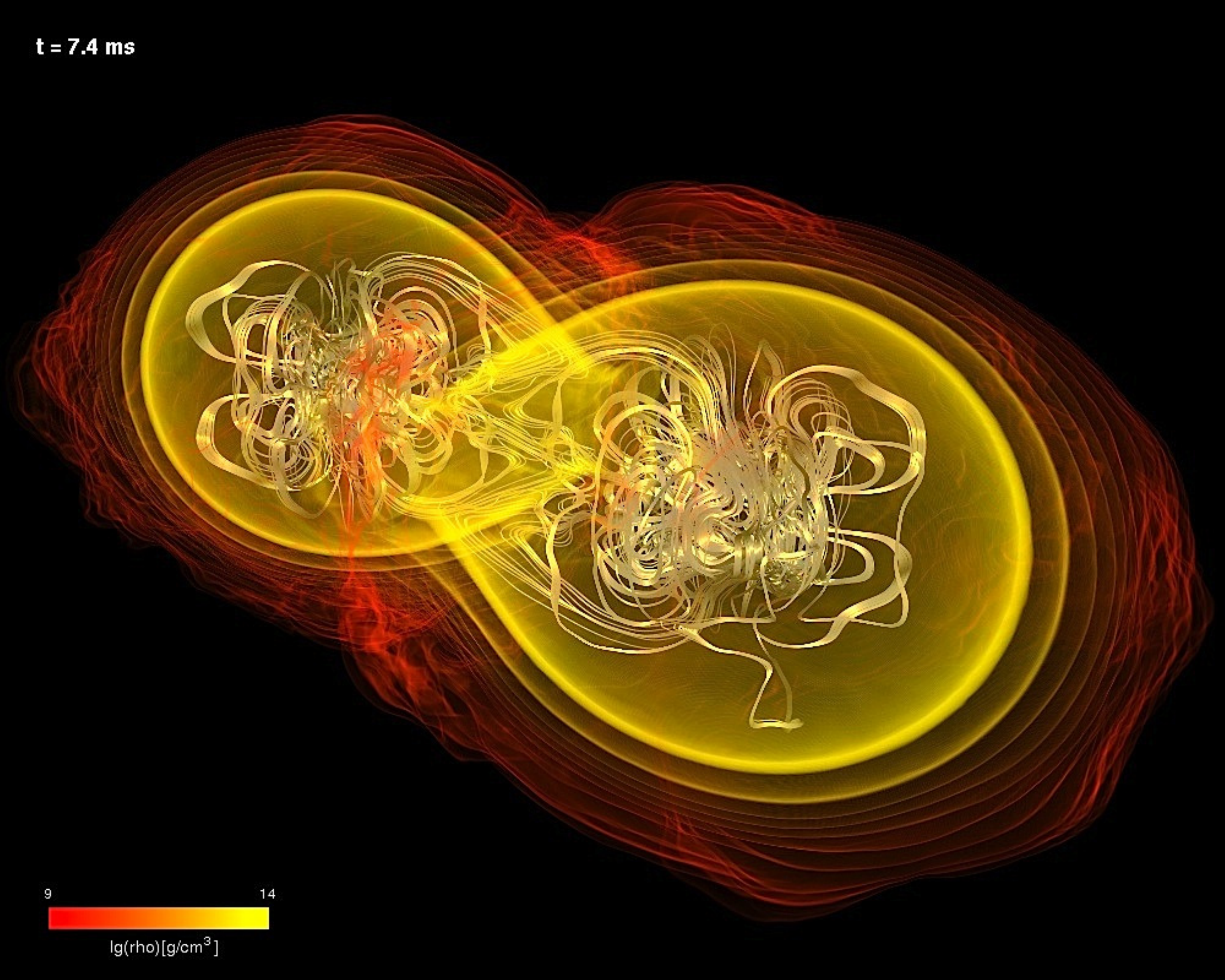}
     \hskip 0.1cm
     \includegraphics[angle=0,width=5.5cm]{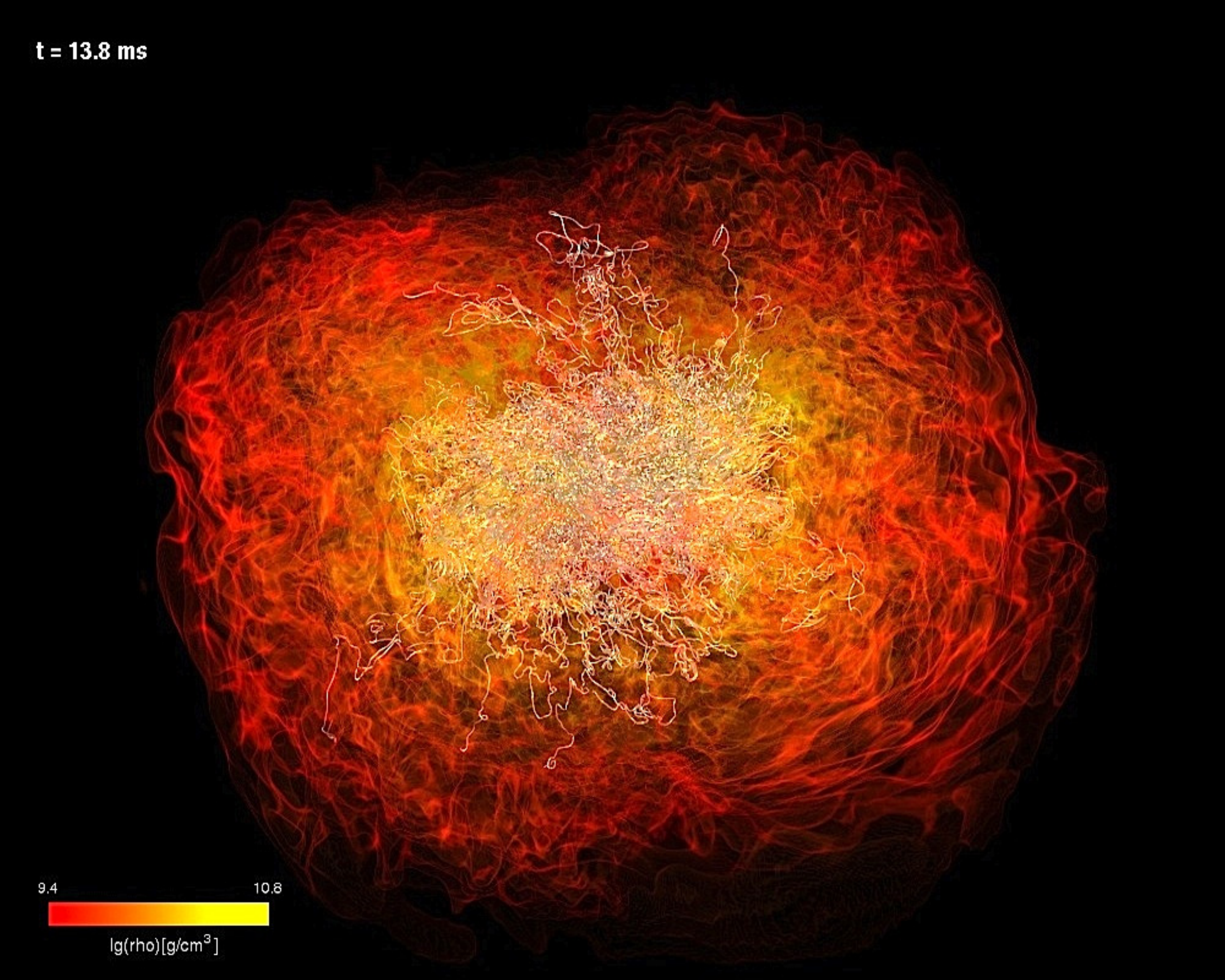}
     \vskip 0.1cm
     \includegraphics[angle=0,width=5.5cm]{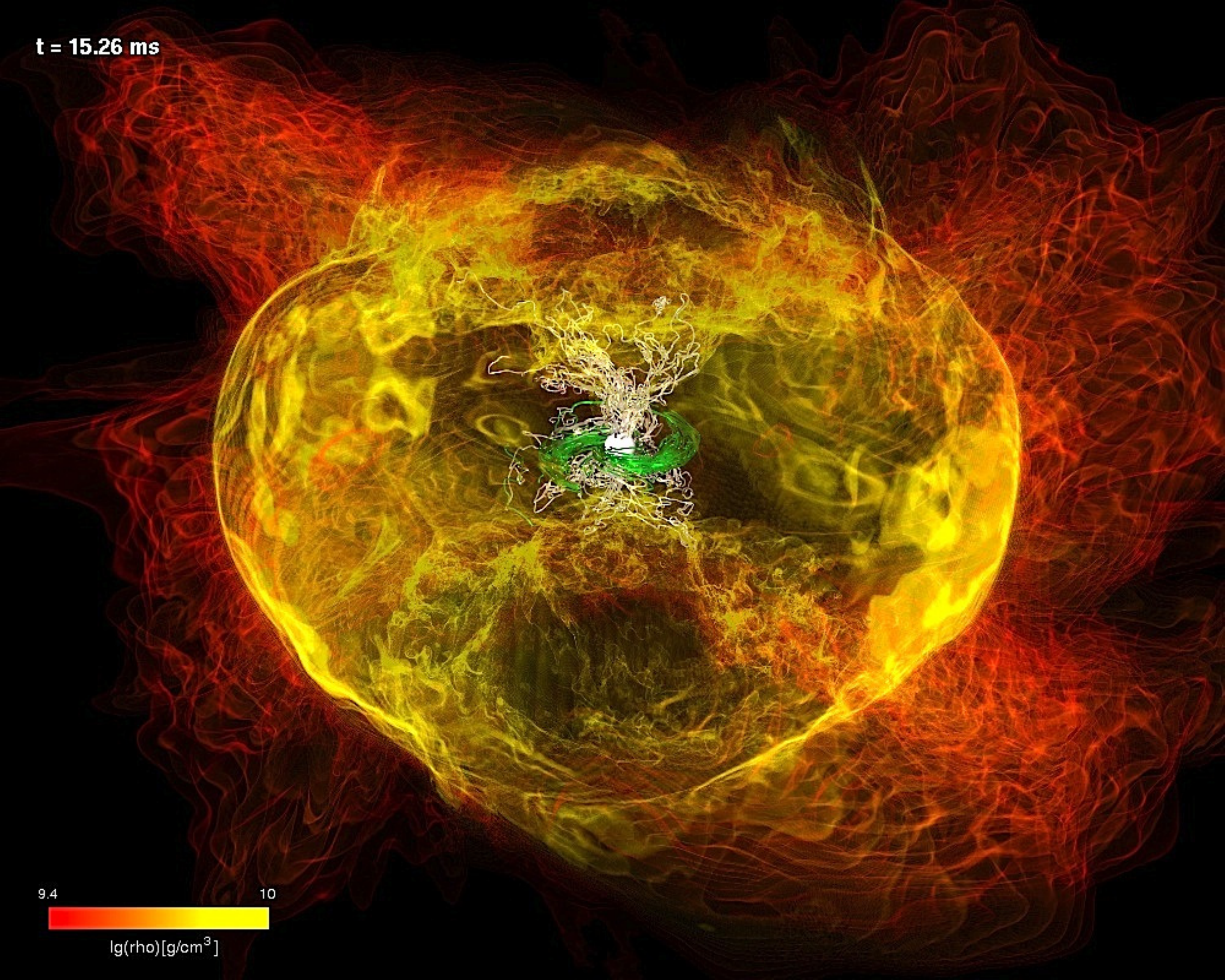}
     \hskip 0.1cm
     \includegraphics[angle=0,width=5.5cm]{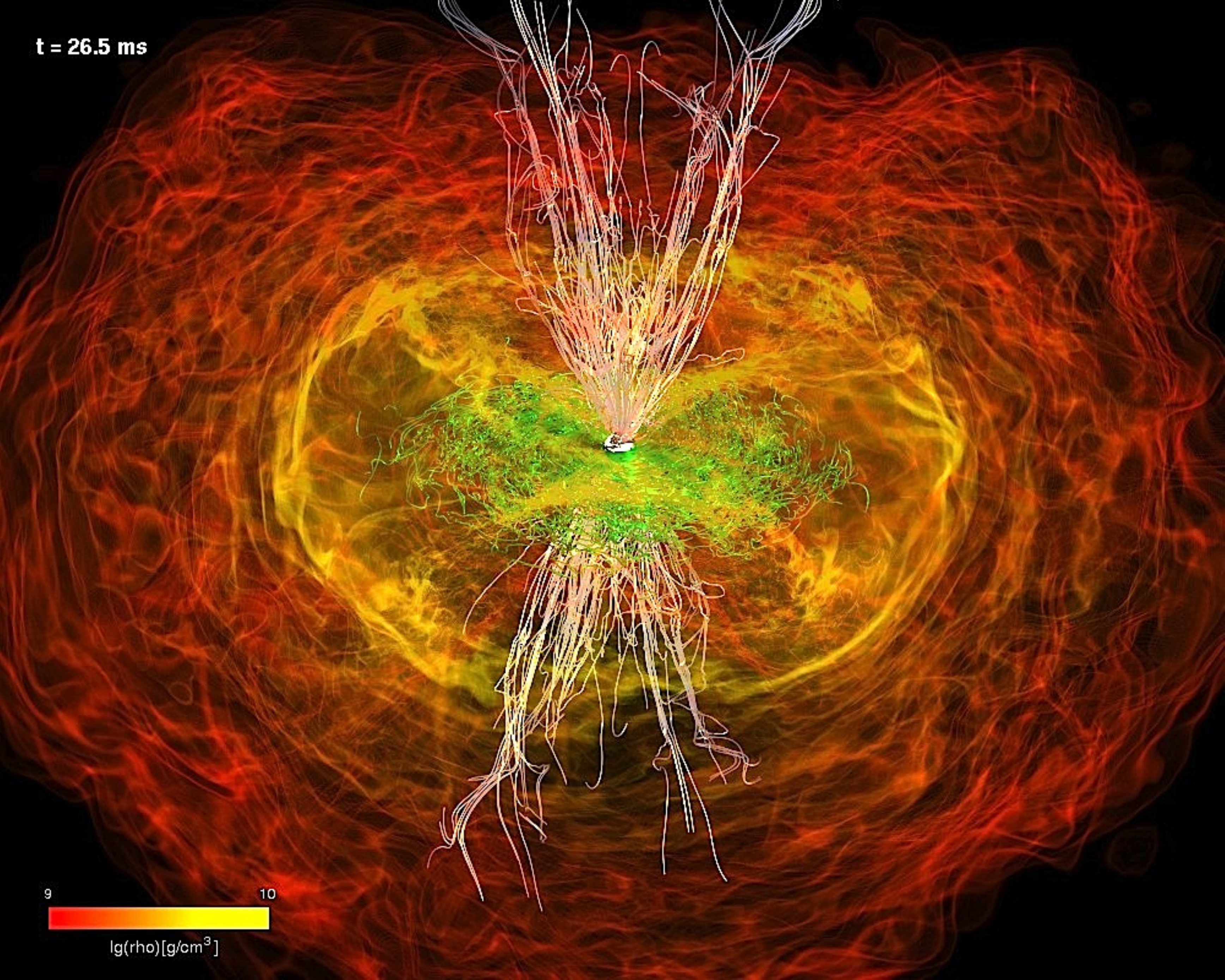}
     \newcapt{Snapshots at representative times of the evolution of the
       binary and of the formation of a large-scale ordered magnetic
       field. Shown with a colour-code map is the density, over which the
       magnetic-field lines are superposed. The panels in the upper row
       refer to the binary during the merger ($t=7.4\,\ms$) and
       \textit{before} the collapse to black hole ($t=13.8\,\ms$), while those in
       the lower row to the evolution \textit{after} the formation of the
       black hole ($t=15.26\,\ms$, $t=26.5\,\ms$). Green lines sample the
       magnetic field in the torus and on the equatorial plane, while
       white lines show the magnetic field outside the torus and near the
       black hole spin axis. The inner/outer part of the torus has a size of
       $\sim90/170\,{\rm km}$, while the horizon has a diameter of
       $\simeq 9\,{\rm km}$.}
  \label{fig:fig1}
\end{center}
\end{figure*}

These stages of the evolution can be seen in Fig.~\ref{fig:fig1}, which
shows snapshots of the density colour-coded between $10^9$ and
$10^{10}\,{\rm gr/cm}^3$, and of the magnetic field lines (green on the
equatorial plane and white outside the torus). Soon after the black hole
formation the torus reaches a quasi-stationary regime, during which the
density has maximum values of $\sim10^{11}\,{\rm g/cm}^3$, while the
accretion rate settles to $\dot{M}\sim0.2\,M_{\odot}/{\rm s}$. Using the
measured values of the torus mass and of the accretion rate, and assuming
the latter will not change significantly, such a regime could last for
$t_{\rm{accr}}\simeq M_{\rm{tor}}/\dot{M}\simeq0.3$~s, after which the
torus is fully accreted; furthermore, if the two neutron stars have
unequal masses, tidal tails are produced which provide additional
late-time accretion~\cite{Rezzolla:2010}. This accretion timescale is
close to the typical observed SGRB
durations~\cite{Kouveliotou1993,Nakar:2007yr}. It is also long enough for
the neutrinos produced in the torus to escape and annihilate in its
neighbourhood; estimates of the associated energy deposition rate range
from $\sim10^{48}\,{\rm erg/s}$~\cite{Dessart2009} to $\sim10^{50}\,{\rm
  erg/s}$~\cite{Setiawan2004}, thus leading to a total energy deposition
between a few $10^{47}\,{\rm erg}$ and a few $10^{49}\,{\rm erg}$ over a
fraction of a second. This energy would be sufficient to launch a
relativistic fireball, but because radiative losses are accounted yet,
the large reservoir of thermal energy in the torus cannot be extracted in
these simulations.

\begin{figure*}
\begin{center}
  \includegraphics[angle=0,width=5.5cm]{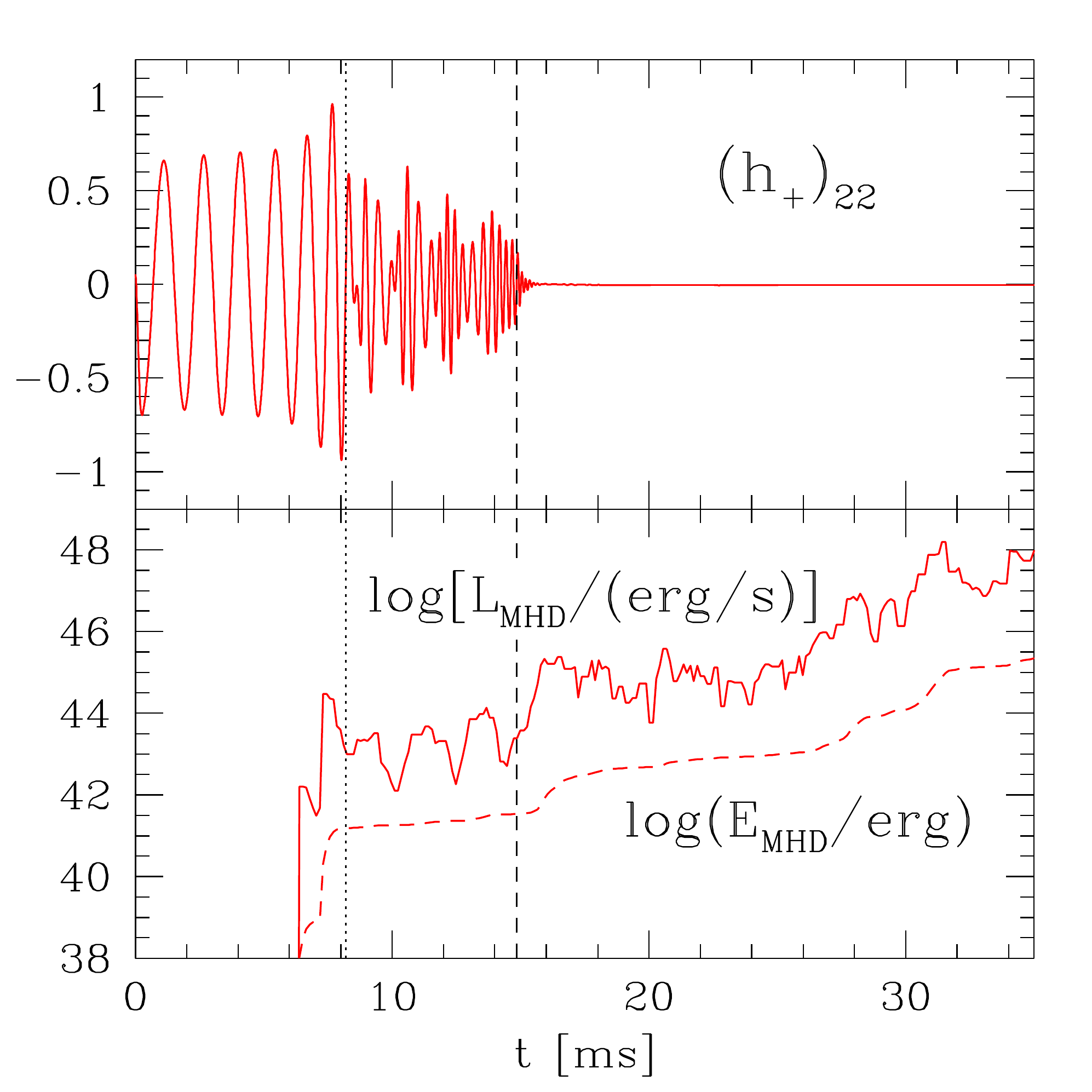}
  \hskip 0.1cm
  \includegraphics[angle=0,width=5.5cm]{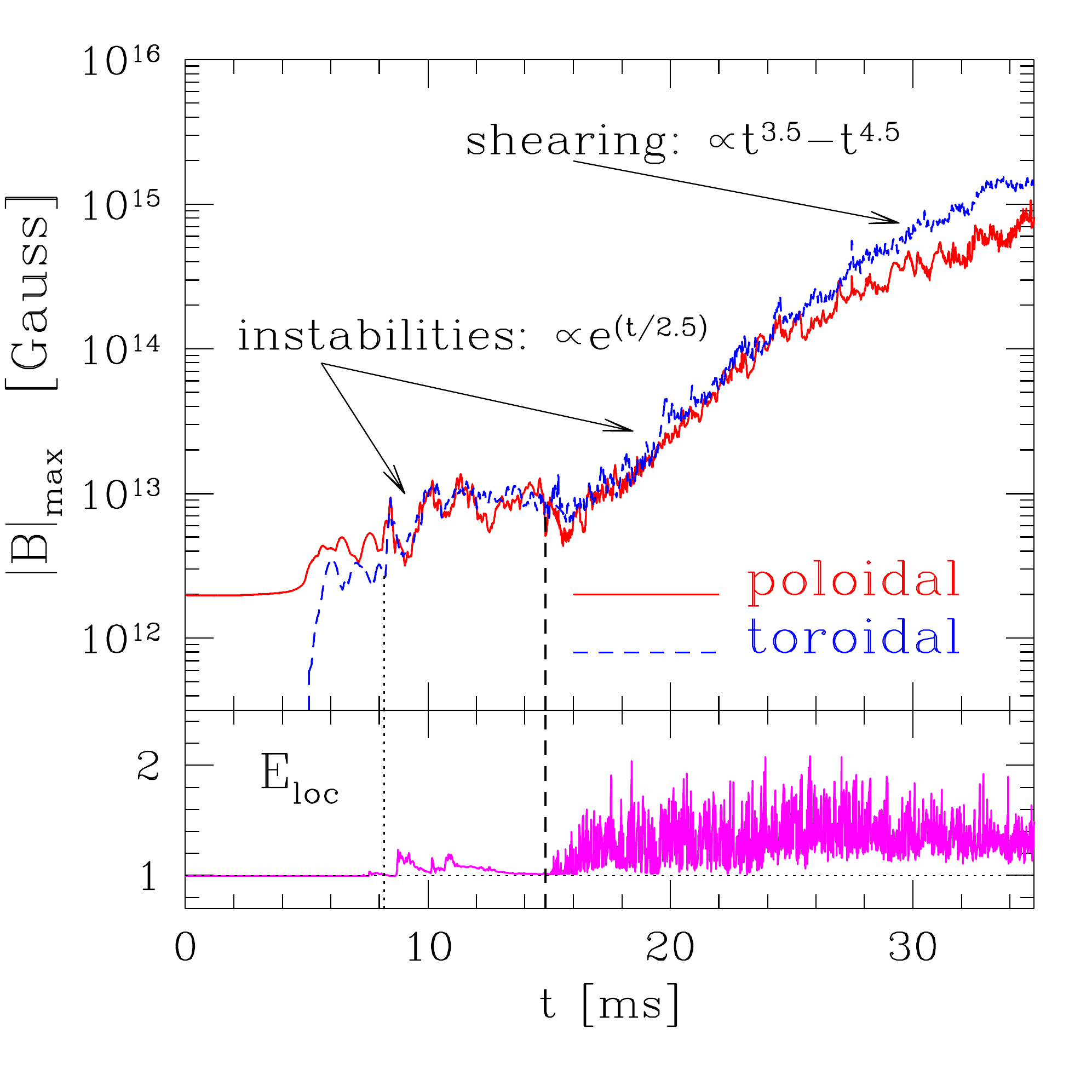}
  \newcapt{\textit{Left panel:} gravitational wave signal shown through
    the $\ell=2,m=2$ mode of the $+$ polarization, $(h_+)_{22}$, (top
    part) and of the MHD luminosity, $L_{\rm MHD}$, (bottom part) as
    computed from the integrated Poynting flux and shown with a solid
    line. The corresponding energy, $E_{\rm MHD}$, is shown with a dashed
    line. The dotted and dashed vertical lines show the times of merger
    (as deduced from the first peak in the evolution of the gravitational
    wave amplitude) and black-hole formation, respectively. \textit{Right
      panel:} Evolution of the maximum of the magnetic field in its
    poloidal (red solid line) and toroidal (blue dashed line)
    components. The bottom panel shows the maximum local fluid energy
    indicating that an unbound outflow (\ie $E_{\rm{loc}} > 1$) develops
    and is sustained after black-hole formation.}
  \label{fig:fig2}
\end{center}
\end{figure*}

The gravitational wave signal of the whole process is shown in the top
part of the left panel in Fig.~\ref{fig:fig2}, while the bottom part
exhibits the evolution of the MHD luminosity, $L_{\rm MHD}$, as computed
from the integrated Poynting flux (solid line) and of the corresponding
energy, $E_{\rm MHD}$, (dashed line). Clearly, the MHD emission starts
only at the time of merger and increases exponentially after black-hole
formation, when the gravitational wave signal essentially shuts
off. Assuming that the quasi-stationary MHD luminosity is
$\sim4\times10^{48}\,{\rm erg/s}$, the total MHD energy released during
the lifetime of the torus is $\sim1.2\times10^{48}\,{\rm erg}$, which, if
spread over an opening half-angle of $\sim30^\circ$ (see discussion
below), suggests a lower limit to the isotropic equivalent energy in the
outflow of $\sim9\times10^{48}\,{\rm erg}$. While this is at the low end
of the observed distribution of gamma-ray energies for SGRBs, larger MHD
luminosities are expected either through the additional growth of the
magnetic field via the winding of the field lines in the
differentially-rotating disk (the simulation covers only one tenth of
$t_{\rm accr}$), or when magnetic reconnection (which cannot take place
within an ideal-MHD approach), is also accounted for [which may also
  increase the gamma-ray efficiency, \eg \cite{McKinney2010}].

The last two panels of Fig.~\ref{fig:fig1} offer views of the accreting
torus after the black-hole formation. Although the \textit{matter}
dynamics is quasi-stationary, the last two panels clearly show that the
\textit{magnetic-field} is not and instead evolves significantly. It is
only when the system is followed well after the formation of a black
hole, that MHD instabilities develop and generate the central,
low-density, poloidal-field funnel. This regime, which was not accessible
to previous simulations~\cite{Price06, Anderson2008, Liu:2008xy}, is
essential for the jet formation~\cite{Aloy:2005,
  Komissarov:2009}. Because the strongly magnetised matter in the torus
is highly conductive, it shears the magnetic-field lines via differential
rotation. A measurement of the angular-velocity in the torus indicates
that it is essentially Keplerian and thus unstable to the
magneto-rotational instability \cite{BalbusHawley1998}, which develops
$\simeq5\,\ms$ after black-hole formation and amplifies exponentially
both the poloidal and the toroidal magnetic fields; the e-folding time of
the instability is $\simeq2.5\,\ms$ and in good agreement with the one
expected in the outer parts of the torus~\cite{BalbusHawley1998}. Because
of this exponential growth, the final value of the magnetic field is
largely insensitive to the initial strength and thus a robust feature of
the dynamics (see also~\cite{Siegel2013} for a similar behaviour recently
computed in a HMNS)

A quantitative view of the magnetic-field growth is shown in the right
panel of Fig.~\ref{fig:fig2}, which shows the evolution of the maximum
values in the poloidal and toroidal components. Note that the latter is
negligibly small before the merger, reaches equipartition with the
poloidal field as a result of a Kelvin-Helmholtz instability triggered by
the shearing of the stellar surfaces at
merger~\cite{Price06,Giacomazzo:2010}, and finally grows to
$\simeq10^{15}$ G by the end of the simulation. At later times
($t\gtrsim22\,\ms$), when the instability is suppressed, the further
growth of the field is due to the shearing of the field lines and it
increases only as a power-law with exponent $3.5~(4.5)$ for the
poloidal~(toroidal) component. Although the magnetic-field growth
essentially stalls after $t\simeq35\,\ms$, further slower growths are
possible~\cite{Obergaulinger:2009}, yielding correspondingly larger
Poynting fluxes. Indeed, when the ratio between the magnetic flux across
the horizon and the mass accretion rate becomes sufficiently large, a
Blandford-Znajek mechanism~\cite{Blandford1977} may be
ignited~\cite{Komissarov:2009dn}; such conditions are not met over the
timescale of the simulations, but could develop over longer
timescales. Also shown in the right panel of Fig.~\ref{fig:fig2} is the
maximum local fluid energy, highlighting that an \textit{unbound outflow}
(\ie $E_{\rm{loc}}>1$) develops after black-hole formation along the
outer walls of the torus and persists for the whole duration of the
simulation.

\begin{figure*}
\begin{center}
  \includegraphics[angle=0,width=5.5cm]{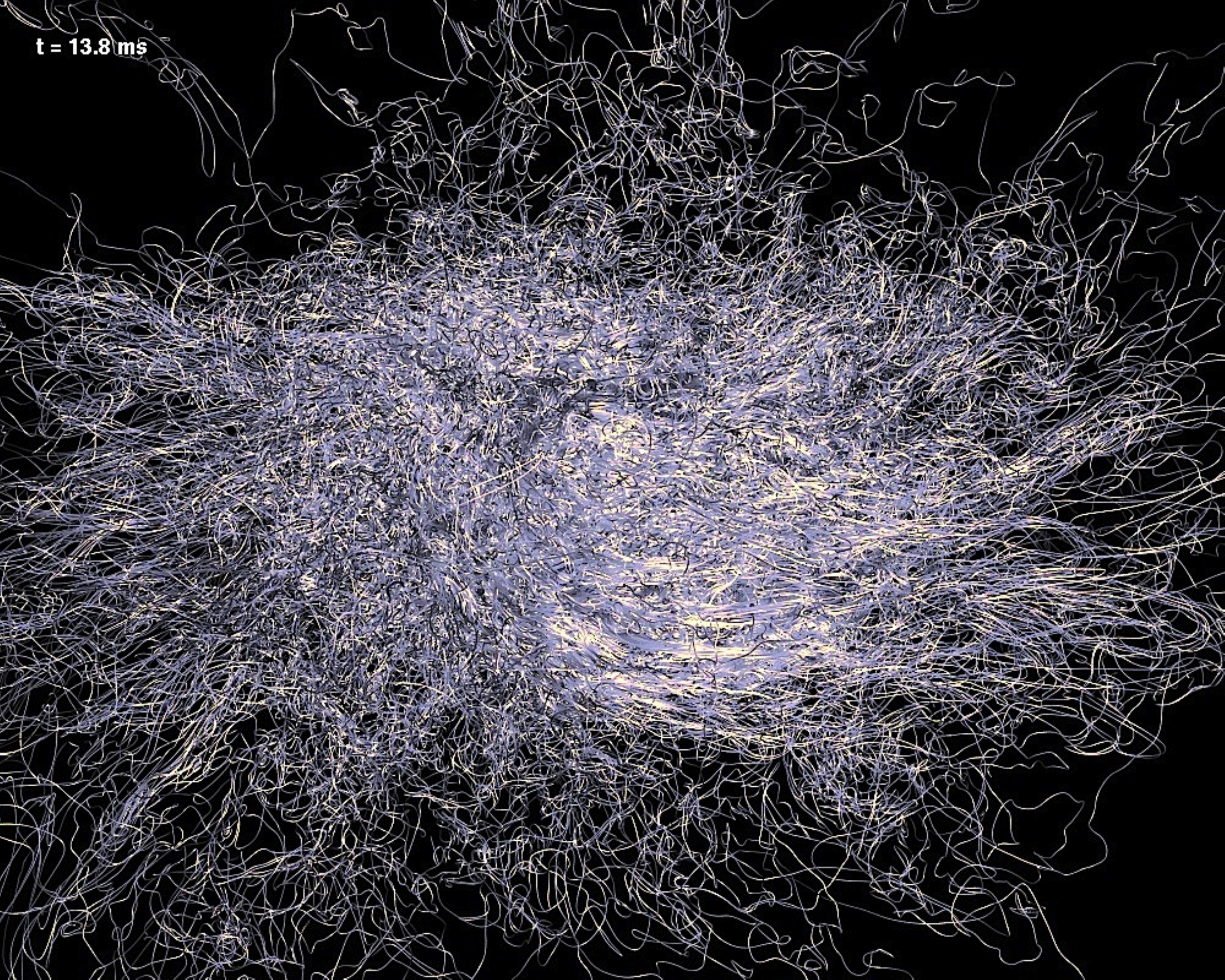}
  \hskip 0.1cm
  \includegraphics[angle=0,width=5.5cm]{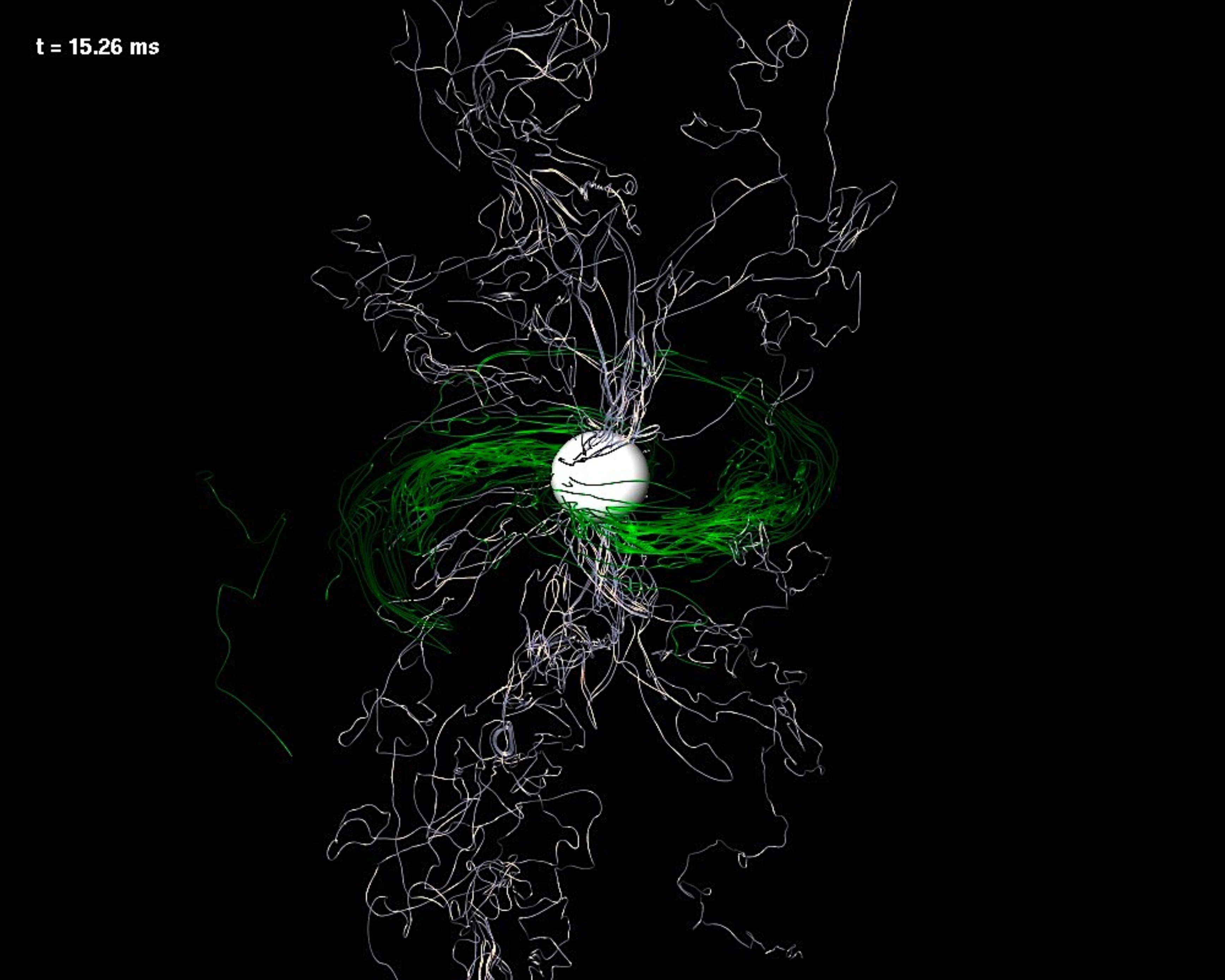}
  \vskip 0.1cm
  \includegraphics[angle=0,width=5.5cm]{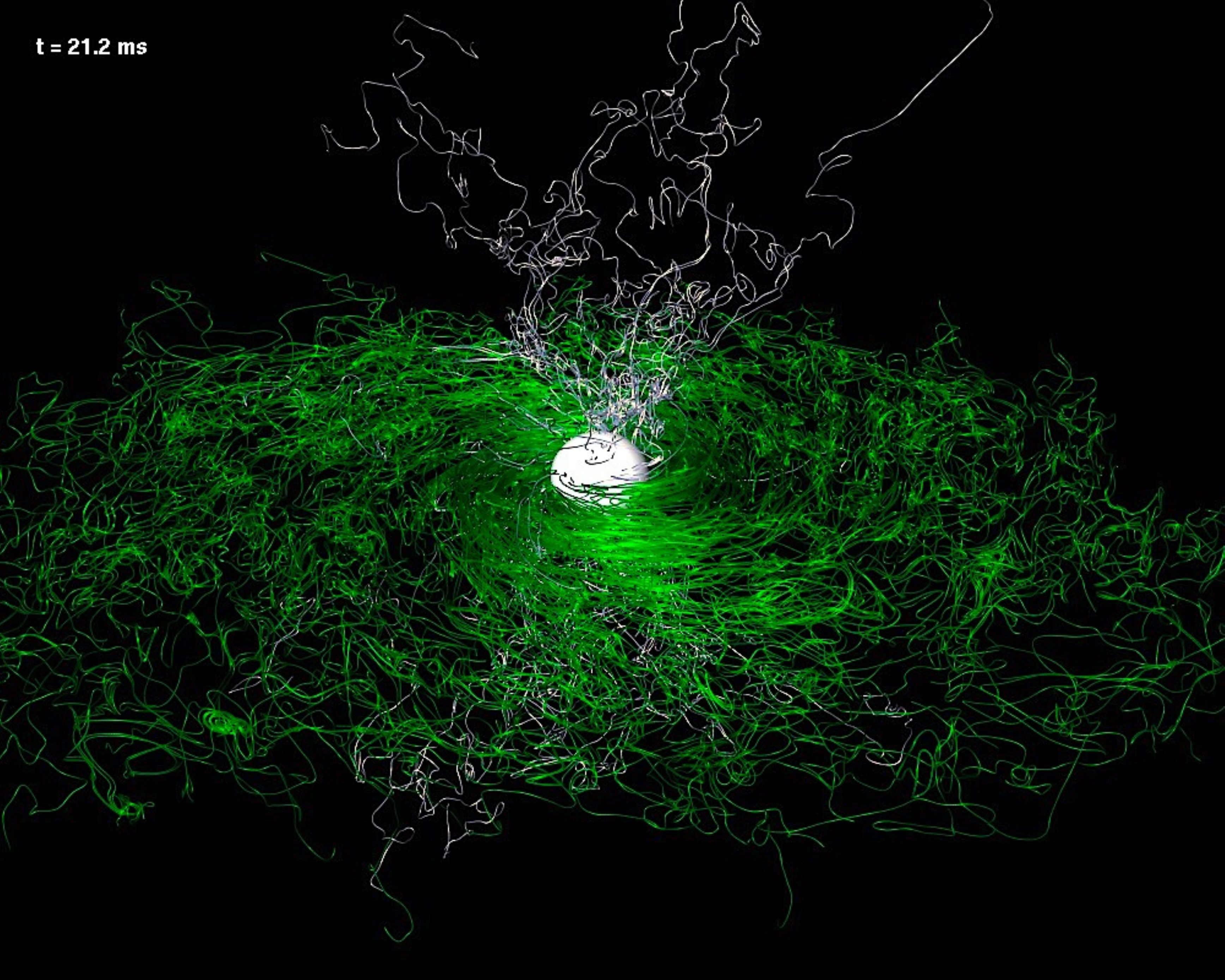}
  \hskip 0.1cm
  \includegraphics[angle=0,width=5.5cm]{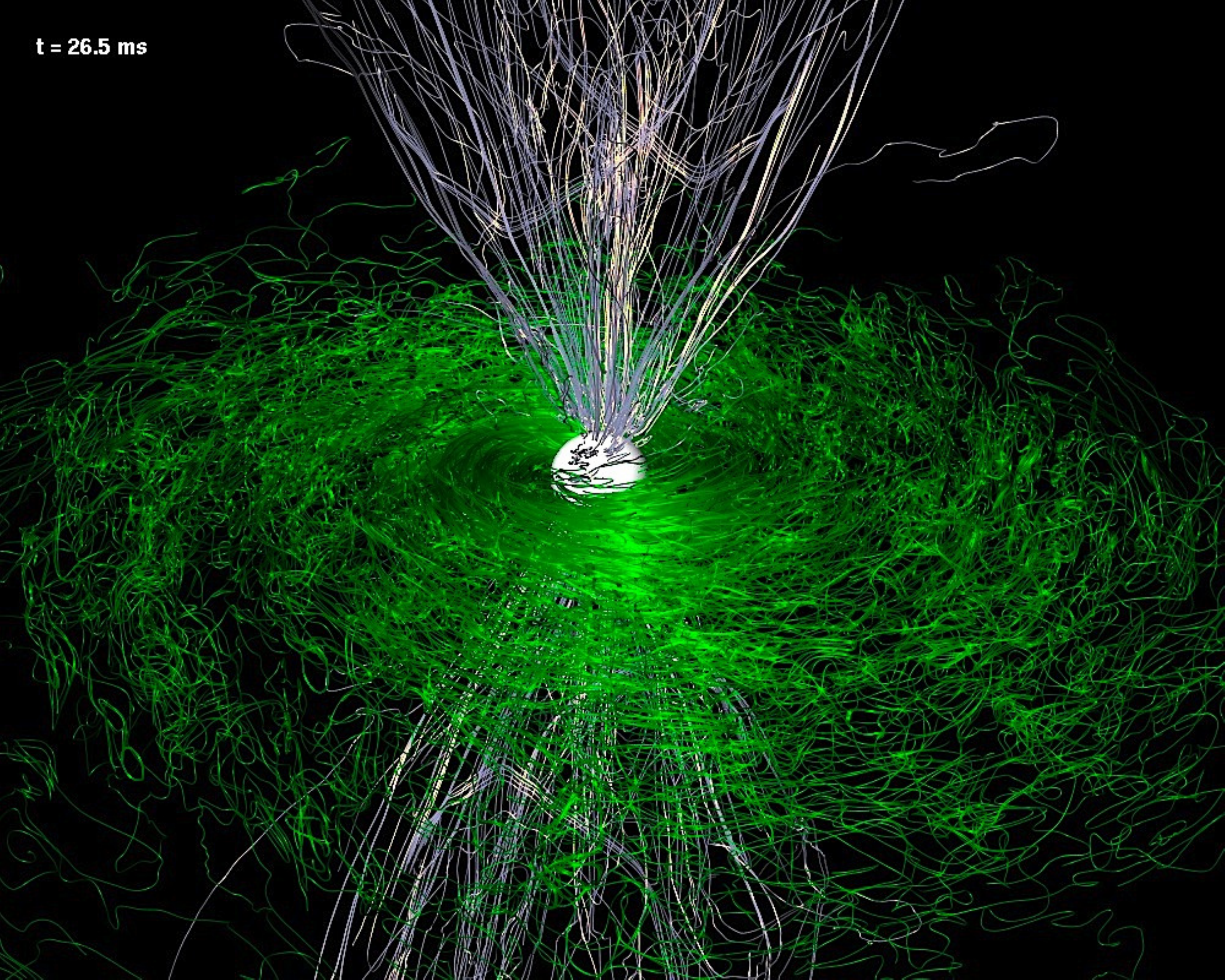}
  \newcapt{Magnetic-field structure in the HMNS (first panel) and after
    the collapse to black hole (last three panels). Green refers to
    magnetic-field lines inside the torus and on the equatorial plane,
    while white refers to magnetic-field lines outside the torus and near
    the axis. The highly turbulent, predominantly poloidal magnetic-field
    structure in the HMNS ($t=13.8\,\ms$) changes systematically as the
    black hole is produced ($t=15.26\,\ms$), leading to the formation of a
    predominantly toroidal magnetic field in the torus
    ($t=21.2\,\ms$). All panels have the same linear scale, with the
    horizon's diameter being of $\simeq 9\,{\rm km}$.}
  \label{fig:fig3}
\end{center}
\end{figure*}

Finally, Fig.~\ref{fig:fig3} provides a summary of the magnetic-field
dynamics. It shows the magnetic field in the HMNS formed after the merger
and its structure and dynamics after the collapse to black hole. In
particular, in the last three panels it shows the magnetic-field
structure inside the torus and on the equatorial plane (green), and
outside the torus and near the axis (white). It is apparent that the
highly turbulent magnetic field in the HMNS ($t=13.8\,\ms$) changes
systematically as the black hole is produced ($t=15.26\,\ms$), leading to
the formation of a toroidal magnetic field in the torus
($t=21.2\,\ms$)\newfoot{Turbulence in relativistic flows is an extremely
  challenging problem that is also essentially unexplored. Also in this
  case, the first relativistic simulations have been performed only
  recently~\cite{Radice2012b,Zrake2013}.}. As the MRI sets in, the
magnetic field is not only amplified, but also organises itself into a
dual structure, which is mostly toroidal in the accretion torus with
$B_{\rm tor}\sim2\times10^{15}\,{\rm G}$, but predominantly poloidal and
jet-like along the black-hole spin axis, with $B_{\rm
  pol}\sim8\times10^{14}\,{\rm G}$ ($t=26.5\,\ms$)\newfoot{A similar
  magnetic-field configuration has been recently reproduced also when
  simulating the merger of a magnetised neutron star onto a black
  hole~\cite{Etienne2012b}.}. Note that the generation of an ordered
large-scale field is far from trivial and a nonlinear dynamo may explain
why the MRI brings a magnetic field self-organization, as it has been
also suggested in case of MRI-mediated growth of the magnetic field in
the conditions met in the collapse of massive stellar
cores~\cite{Lesur:2008, Obergaulinger:2009}. However, the jet-like
structure produced in the simulation is not yet the highly collimated
ultrarelativistic outflow expected in SGRBs (see also below).

The magnetic hollow jet-like structure has an opening half-angle of
$\sim30^\circ$, which sets an upper limit for the opening half-angle of
any potential outflow, either produced by neutrino energy
deposition~\cite{Aloy:2005} or by electromagnetic
processes~\cite{Komissarov:2009}. In these simulations most of the
outflow develops along the edges of the jet-like structure, via a
turbulent layer of electromagnetic driven matter, which shields the central funnel
from excessive baryonic pollution. It is reasonable to expect that such a
layer is crucial to set the opening angle of any ultrarelativistic jet,
to shape both the radial and transverse structure of the jet, as well as
to determine its stability properties. The Lorentz factors of the outflow
measured in these simulations are not very high ($\gamma \lesssim4$), but
can potentially be amplified by several orders of magnitude in the inner
baryon-poor regions through special-relativistic
effects~\cite{Aloy:2006rd}, the variability of the
flow~\cite{Granot2010}, or when resistive-MHD effects are taken into
account~\cite{Dionysopoulou:2012}. Such accelerations will be produced as
a more realistic and general-relativistic treatment of the radiative
losses will become computationally affordable.

\subsection{Comparison with observations}

Below I briefly discuss how the results presented above broadly match the
properties of the central engine as deduced from the observations.

\smallskip
\noindent\textit{Duration}: The observed duration of the prompt
gamma-ray emission GRBs is energy dependent and is usually determined
through $T_x$, the time over which $x\%$ of the total counts are
observed, between the $(100-x)/2$ and $(100+x)/2$ percentiles. The most
common intervals used are $T_{90}$ (or $T_{50}$), initially
defined~\cite{Kouveliotou1993} between $20\;$keV and $2\;$MeV. The GRB
duration distribution is bimodal~\cite{Kouveliotou1993}, where the
durations of SGRBs (approximately 25\% of GRBs) are well-fit by a fairly
wide log-normal distribution centred around $T_{90}\approx 0.8\;$s with
a FWHM of 1.4 dex~\cite{Nakar:2007yr}. The typical redshifts of the SGRBs
observed with {\it Swift} are in the range $z \sim0.3-1$, suggesting a
central value of the intrinsic duration distribution of
$\approx\langle1+z\rangle^{-1}0.8\;{\rm s}\sim0.5\;$s, and a comparably
wide distribution around this value. This is in close agreement with our
accretion time of $\sim0.3\;$s.

\smallskip
\noindent\textit{Energy}: The isotropic equivalent energy output in the
prompt gamma-ray emission of SGRBs, $E_{\rm\gamma,iso}$, spans a wide
range, from $(2.7\pm1)\times 10^{48}\;$erg (in the observed energy range
$15-350\;$keV) for GRB~050509B at a redshift of $z =
0.225$~\cite{Bloom2006}, up to $(1.08\pm0.06)\times 10^{53}\;$erg (in the
observed energy range 10$\;$keV -- 30$\;$GeV) for GRB~090510 at
$z=0.903$~\cite{Ackermannn2010}. However, the most typical values are in
the range $E_{\rm\gamma,iso} \sim 10^{49} -
10^{51}\;$erg~\cite{Nakar:2007yr}. In this model, the highly relativistic
outflow may be powered either by neutrino-anti neutrino annihilation, or
by the Blandford-Znajek mechanism. For the former one might expect a
total energy release between a few $10^{47}\;$erg and
$\sim10^{49}\;$erg~\cite{Oechslin06,Birkl2007}, into a bipolar
relativistic jet of opening half-angle $\theta_{\rm jet}\sim8-30^\circ$,
corresponding to a fraction $f_b\sim0.01-0.13$ of the total solid angle,
and isotropic equivalent energies, $E_{\rm \nu\bar{\nu},iso}$, between a
few $10^{48}\;$erg and $\sim10^{51}\;$erg. For the latter mechanism,
instead, and if the magnetisation near the event horizon becomes
sufficiently high, the jet power for these values for the black-hole mass
and spin is~\cite{Lee2000}
\begin{equation}
L_{\rm BZ} \sim3.0\times10^{50}\left(\frac{f_{\rm rel}}{0.1}\right)
\left(\frac{B}{2\times10^{15}\,\rm{G}}\right)^2\, {\rm erg/s}\,,
\end{equation}
where $f_{\rm rel}$ is the fraction of the total Blandford-Znajek power
that is channelled into the resulting relativistic jet (and $f_{\rm
  rel}\sim0.1$ might be expected for ejecta with asymptotic Lorentz
factors above $100$). This relativistic outflow is launched over a
timescale of $\sim0.2\,{\rm s}$ and corresponds to
\begin{equation}
E_{\rm BZ, iso}\sim1.2\times10^{51} \left(\frac{f_{\rm rel}}{0.1}\right)
\left(\frac{f_b}{0.05}\right)^{-1}
\left(\frac{B}{2\times 10^{15}\,\rm{G}}\right)^2\,{\rm erg}\,,
\end{equation}
Comparing the X-ray afterglow luminosity (after 10 or 11 hours) and
$E_{\rm\gamma,iso}$ suggests that the efficiency of the prompt gamma-ray
emission in SGRBs is typically high ~\cite{Bloom2006}, and similar to
that of long GRBs ~\cite{Granot2006}, with
$E_{\rm\gamma,iso}\sim(0.1-0.9)E_{\rm iso}$, radiating between $\sim10\%$
and $\sim90\%$ of the initial energy of the ultrarelativistic
outflow. Therefore, this model is able to accommodate the observed
$E_{\rm\gamma,iso}$ values.

\smallskip
\noindent\textit{Lorentz factor}: The Fermi Gamma-Ray Space Telescope has
detected GeV emission from SGRBs~\cite{Abdo2009a}, suggesting typical
lower limits of $\gamma_{\rm min}\sim10^2-10^3$. In particular,
$\gamma_{\rm min}\approx1200$ was obtained for
GRB~090510~\cite{Ackermannn2010}. However, a more realistic model
~\cite{Granot2008} results in $\gamma_{\rm min}$ values lower by a factor
of $\sim 3$. Therefore, the central engine should be capable of producing
outflow Lorentz factors of at least a few hundred. The fact that our
simulation produces a strongly magnetised mildly relativistic outflow at
angles near $\sim30^\circ$ from the black-hole spin axis would help
shield the inner region near the spin axis from excessive baryon loading,
and thus assist in achieving high asymptotic Lorentz factors at large
distance from the source, after the outflow in this region is triggered
by neutrinos and/or the Blandford-Znajek mechanism.

\smallskip
\noindent\textit{Jet angular structure}: This is poorly constrained
by observations (even more so than for long GRBs). The only compelling
case for a jet break in the afterglow light-curve is for
GRB~090510~\cite{DePasquale:2010}, which occurred very early on (after
$\sim1400\;$s), and would thus imply an extremely narrow jet
($\theta_{\rm jet}\sim0.2-0.4^\circ$) and modest true energy output in
gamma-rays ($\sim10^{48}\;$erg). If this is indeed a jet break, it might
correspond to a line of sight near a very narrow and bright core of a
jet, which also has significantly wider wings. Observers with lines of
sight along these wings would then see a much dimmer and more typical
SGRB~\cite{Rossi2002,Peng2005}; without such wings, however, the
observations would suggest a very large intrinsic and beaming-corrected
event rate per unit volume. In most cases there are only lower limits on
a possible jet break time~\cite{Nakar:2007yr}, resulting in typical
limits of $f_b \gtrsim 10^{-2}$ or $\theta_{\rm jet} \gtrsim
8^\circ$. This is consistent with the expectation of $\theta_{\rm
  jet}\sim8-30^\circ$ for the ultrarelativistic ejecta capable of
producing a SGRB (which would also imply a reasonable SGRBs intrinsic
event rate per unit volume).

\subsection{Summary}

The calculations reported above demonstrate that a binary merger of two
neutron stars inevitably leads to the formation of a relativistic
jet-like and ultrastrong magnetic field, which could serve as a central
engine for SGRBs. Because the magnetic-field growth is exponential, the
picture emerging from these simulations is rather general and applies
equally even to mildly magnetised neutron stars. Overall, this first
``little piece'' of numerical relativity removes a significant uncertainty as to
whether such binary mergers can indeed produce the central engines of
SGRBs. While the electromagnetic energy release is already broadly
compatible with the observations, the simulations discussed above lack a
proper treatment of the energy losses via photons and neutrinos or
resistive dissipation, which can provide a fundamental contribution to
the energy-input necessary to launch the fireball and cool the
torus~\cite{Setiawan2004,Dessart2009}. This additional energy input,
whose self-consistent inclusion in general relativity remains extremely
challenging, may help to launch an ultrarelativistic outflow very early
after the black hole forms and complete the picture of the central engine
of a SGRB.

\section{Second piece: A dynamical hoop conjecture}

The second ``little piece'' of numerical relativity that I will discuss
aims at addressing the issue of necessary conditions for the formation of
a black hole, which still represents one of the most intriguing and
fascinating predictions of classical general relativity. There is
abundant astronomical evidence that black holes exist, and a number of
considerations supporting the idea that under suitable conditions
gravitational collapse is inevitable~\cite{Wald84}. In addition, there is
overwhelming numerical evidence that black-hole formation does take place
in a variety of environments~\cite{Rezzolla:2011}. Yet, a rigorous
definition of the sufficient conditions for black-hole formation is still
lacking. Hence, it is not possible to predict whether the collision of
two compact objects, either stars or elementary particles, will lead to
the formation of a black hole.

The \textit{hoop conjecture} proposed by Thorne in the '70s, provides
some reasonable and intuitive guidelines~\cite{Thorne72a}. I recall that
the conjecture states that a black hole is formed if an amount of
``mass-energy'' $E$ can be compressed to fit within a hoop with radius
equal or smaller than the corresponding Schwarzschild radius, \ie if
\hbox{$R_{\mathrm{hoop}} \leq R_\mathrm{s} = 2GE/c^4$}, where $G$ is
gravitational constant and $c$ the speed of light. Even though it can be
made precise under particular circumstances~\cite{senovilla_2008_rhc},
the hoop conjecture is not meant to be a precise mathematical statement
and, in fact, it is difficult to predict if the above-mentioned collision
will compress matter sufficiently to fit within the limiting
hoop. Loosely speaking, what is difficult is to determine which part of
the ``kinetic energy'' of the system can be accounted to fit within the
hoop. Since at the collision the conversion of kinetic energy into
internal energy is a highly nonlinear process, any quantitative
prediction becomes rapidly inaccurate as the speeds involved approach
that of light.

As stated above, the hoop conjecture is purely classical. A
quantum-mechanical equivalent is not difficult to formulate, although not
very stringent, as it simply implies that a black hole will be formed at
Planck-energy scales. The predicting power does not improve significantly
when considering the conditions of black-hole formation in
higher-dimensional theories of gravity (see, \eg
\cite{Argyres_1998,Yoshino_2003,Yoo_2010}). In these frameworks, the
energy required for black-hole formation might be significantly
smaller~\cite{Argyres_1998}, thus providing the possibility of producing
them in the Large Hadron Collider (LHC)~\cite{Dimopoulos_2001}, but no
firm conclusion has been reached yet.

Clearly, although numerical simulations represent a realistic route to
shed some light on this issue (see, \eg,
\cite{Sperhake:2008ga,Shibata2008,Sperhake2009}), even the simplest
scenario of the collision of two compact objects at ultrarelativistic
speeds is far from being simple and it is actually very challenging. A
first step was taken by Eardley and Giddings~\cite{Eardley_2002}, who
have studied the formation of a black hole from the head-on collision of
two plane-fronted gravitational waves with nonzero impact parameter
(previous work of D'Eath and
Payne~\cite{Death1992a,Death1992b,Death1992c} using different methods had
considered a zero impact parameter). In all of these analyses each
incoming particle is modelled as a point particle accompanied by a
plane-fronted gravitational shock wave corresponding to the
Lorentz-contracted longitudinal gravitational field of the particle. At
the instant of collision the two shock waves pass through one another and
interact through a nonlinear focusing and shearing. As a result of their
investigation, a lower bound was set on the cross-section for black-hole
production, \ie $\sigma > 32.5(GE/2c^4)^2$, where $E$ is the
centre-of-mass (lab) energy. More recently, and in a framework which is
closer to the one considered here, this problem has been investigated by
Choptuik and Pretorius~\cite{Choptuik:2010a}, who studied the collision
of two classical spherical solitons, with a total energy of the system in
the lab frame $E=2 \gamma_{\mathrm{b}} m_0 c^2$, where $m_0$ is the
``rest-mass'', $\gamma_{\mathrm{b}} \equiv 1/\sqrt{1-
  v^2_{\mathrm{b}}/c^2}$ and $v_{\mathrm{b}}$ the boost velocity. They
were then able to show that for collisions with sufficiently high boost,
\ie $\gamma_{\mathrm b} \gtrsim 2.9$, a black hole can be formed.

In what follows I discuss what has been recently reported on the first
calculations of black-hole production from the collision of two compact,
selfgravitating, fluid objects boosted at ultrarelativistic
speeds\newfoot{Much of what follows is taken from the discussion
  presented in Ref.~\cite{Rezzolla2013}.} (A similar investigation by
East and Pretorius~\cite{East2012} has also appeared at about the same
time).

I start by pointing out that there are several important differences with
the previous investigations
in~\cite{Death1992a,Death1992b,Death1992c,Eardley_2002,Choptuik:2010a}. Differently
from~\cite{Death1992a,Death1992b,Death1992c,Eardley_2002}, in fact, I
will consider colliding objects that are not in vacuum and are not
treated as point particles. Rather, they are relativistic stars, which
obviously extended and selfgravitating objects, thus with a behaviour
that is intrinsically different. Also, differently
from~\cite{Choptuik:2010a}, these objects are not described as scalar
fields, but as fluids and thus represent a more realistic description of
baryonic matter, such as the one employed when simulating relativistic
heavy-ion collisions~\cite{Rischke1995}. These intrinsic differences also
make the comparison with the works
of~\cite{Death1992a,Death1992b,Death1992c,Eardley_2002} very hard if
possible at all. On the other hand, many analogies exist with the
collision of bosons stars considered in~\cite{Choptuik:2010a}, and that,
as I will discuss below, can be interpreted within the more general
description of black-hole production from ultrarelativistic collisions.

Overall, the most important and distinguishing feature in the collision
of two selfgravitating stars is that a black hole can be produced even
from zero initial velocities if the initial masses are large enough; this
behaviour is clearly absent in all previous results, where instead a
critical initial boost is necessary~\cite{Death1992a,Death1992b,
  Death1992c, Eardley_2002, Choptuik:2010a}. In addition, for each value
of the effective Lorentz factor, $\langle \gamma \rangle$, a critical
initial mass exists, $M_{\rm c}$, above which a black hole is formed and
below which matter, at least in part, selfgravitates. More importantly,
both $M_{\rm c}$ follows a simple scaling with $\langle \gamma \rangle$,
thus allowing to extrapolate the results to the masses and energies of
modern particle accelerators and to deduce that black-hole production is
unlikely at LHC scales.

\subsection{The numerical setup}

The numerical setup employed in the simulations is the same presented
in~\cite{Kellermann:10}, and it uses an axisymmetric code to solve in two
spatial dimensions, $(x,z)$, the set of the Einstein and of the
relativistic-hydrodynamic equations~\cite{Kellermann:08a}. The
axisymmetry of the spacetime is imposed exploiting the ``cartoon''
technique, while the hydrodynamics equations are written explicitly in
cylindrical coordinates. All the simulations use an ideal-fluid EOS with
$\Gamma=2$. The initial configurations consist of spherical stars,
constructed as in~\cite{Kellermann:10,Radice:10} after specifying the
central density, $\rho_c$, where the latter also serves as parameter to
determine the critical model. The stars have an initial separation $D$
and are boosted along the $z$-direction via a Lorentz transformation with
boost $\vb$. To limit the initial violation in the constraints, $D$ is
chosen to be sufficiently large, \ie $D=240\,M_{\odot}$, and an optimal
superposition of the two isolated-star solutions that will be presented
in a longer paper. The grid has uniform spacing $\Delta = 0.08
(0.06)\,M_{\odot}$ with extents $x/M_{\odot} \in [0, 80]$ and
$z/M_{\odot} \in [0, 150 (200)]$, where the round brackets refer to the
more demanding high-boost cases. Reflection boundary conditions are
applied on the $z=0$ plane, while outgoing conditions are used elsewhere.

\subsection{The basic dynamics}

The dynamics of the process is rather simple. As the two stars approach
each other, the initial boost velocity increases as a result of the
gravitational attraction, leading to a strong shock as the surfaces of
the stars collide. In the case of \textit{supercritical} initial data,
\ie of stars with masses above a critical value, $M_{\rm c}$, a black
hole is promptly produced and most of the matter is accreted. Conversely,
in the case of \textit{subcritical} initial data, \ie of stars with
masses below $M_{\rm c}$, the product of the collision is a hot and
extended object with large-amplitude oscillations. Part of the stellar
matter is unbound and leaves the numerical grid as the product of the
collision reaches an equilibrium.

\begin{figure*}
\begin{center}
  \includegraphics[width=5.5cm,angle=0]{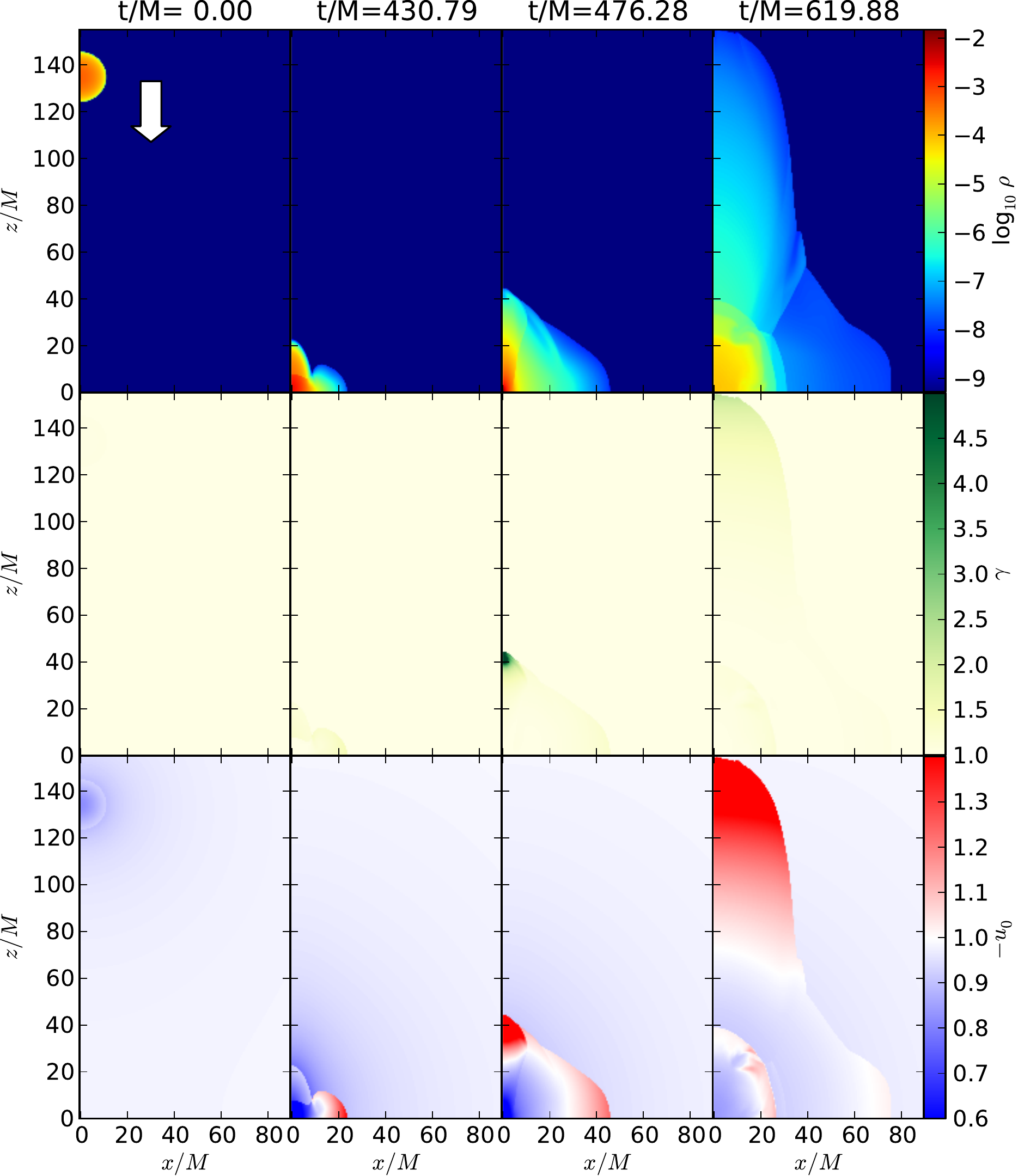}
  \hskip 0.1cm
  \includegraphics[width=5.5cm,angle=0]{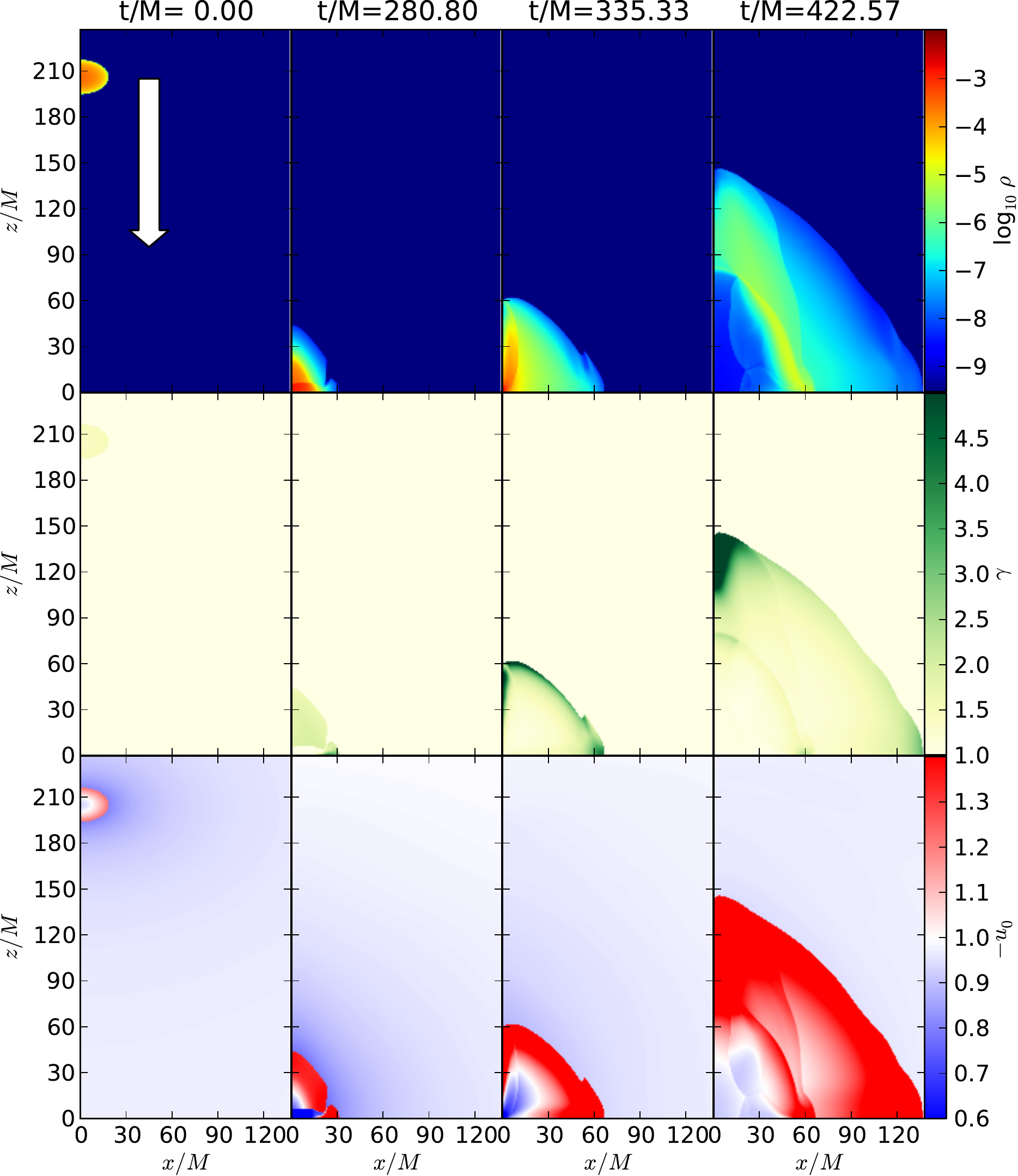}
  \newcapt{Representative snapshots of the rest-mass density, $\rho$ in
    units where $c=1=M_{\odot}$ (top row), of the Lorentz factor,
    $\gamma$ (middle row), and of the local fluid energy, $-u_0$ (bottom
    row), for subcritical models with an initial small boost $\vb=0.3$
    (left panel) or a large one $\vb=0.8$ (right panel). Note that the
    post-collision flow is essentially jet-like for the low-boost case
    (left panel), while essentially spherical for the high-boost case
    (right panel); in this latter case, most of the matter is unbound.}
  \label{fig:fig4}
\end{center}
\end{figure*}

Figure~\ref{fig:fig4} shows snapshots at representative times of the
rest-mass density, $\rho$ (top row), of the Lorentz factor, $\gamma
\equiv (1-v^i v_i/c^2)^{-1/2}$ (middle row), and of the local fluid
energy, $-u_0$ (bottom row), for two subcritical models. The left panel,
in particular, refers to a binary boosted at $\vb=0.3$. Note that the
stars are strongly compressed by the collision, with the rest-mass
density increasing exponentially. The merged object expands in a jet-like
fashion along the $z$-direction, with the bulk of the matter being
accelerated up to $\gamma \sim 16$, or equivalently, $v/c \sim 0.998$,
but then settling on much slower flows with $\gamma \lesssim
2.1$. Furthermore, the front of the jet has $-u_0>1$ indicating that part
of the shocked matter has sufficient energy to have become
gravitationally unbound. As a result, the rest-mass density at the center
of the merged object is smaller than the maximum density of the initial
configuration, although the origin still represents the region where the
density is the largest. The right panel, on the other hand, refers to a
highly-boosted binary, \ie with $\vb=0.8$, with each star being initially
highly distorted by the Lorentz contraction. Also in this case, the stars
are strongly compressed by the collision, but the merged object expands
in a spherical blast-wave fashion, with an almost spherical distribution
of matter and bulk Lorentz factor. The latter reaches values as large as
$\gamma \sim 30$, or equivalently, $v/c \sim 0.999$, which, in contrast
with the low-boost case, do not decrease in time. As a comparison, the
typical bulk Lorentz factors obtained in the merger of binary neutron
stars in quasi circular orbits is $\gamma \sim
1.03$~\cite{Rezzolla:2010}. The very large kinetic energies involved in
the collision are sufficient to make a very large portion of the stellar
matter unbound, as clearly shown by the bottom-right panel of
Fig.~\ref{fig:fig4}, which reports the local fluid energy. The rest-mass
density distribution in the expanding blast wave has a minimum at the
origin, where a large rarefaction is produced by the matter expanding as
an ultrarelativistic thick shell.

The marked transition from a jet-like outflow, not too dissimilar from
the simple Bjorken flow used to model the very early states of
relativistic ion-collisions~\cite{Bjorken1983}, to a shell-like
structure, not too dissimilar from ``transverse expansion'' modelled in
the subsequent stages of relativistic ion-collisions
(see~\cite{Huovinen2006} and references therein), signals that it is not
unreasonable to extrapolate some of the results presented here also to
the collision of ultrarelativistic elementary particles.

The transition from the two qualitatively-different regimes discussed
above is further confirmed by the evolution of the rest-mass normalized
to the initial value $M_0$. The simulations in fact reveal that the
unbound fraction is just a few percent of the total rest-mass in the case
of a low-boost collision, with most of the matter being confined in the
selfgravitating ``star''. This is to be contrasted with what happens for
a high-boost collision, where the unbound fraction is $\sim 100\%$ of the
total rest-mass. This behaviour provides a strong indication that, at
least for subcritical collisions, the role played by gravitational forces
is a minor one as the kinetic energy is increased. This is what happens
in the collision of two particles at ultrarelativistic speeds, where all
of the matter is obviously unbound.

\subsection{Critical behaviour and scaling}

A remarkable property of the head-on collision of compact stars is the
existence of type-I critical behaviour, which was first pointed out
in~\cite{Jin:07a} and subsequently extended in~\cite{Kellermann:08a}. In
essence, in these works it was found that when considering stars with
initial zero velocity at infinity, it is possible to fine-tune the
initial central density $\rho_{\rm c}$ (and hence the mass) near a
critical value, $\rho^{\star}_{\rm c}$, so that stars with $\rho_c >
\rho_c^{\star}$ would collapse \emph{eventually} to a black hole, while
the models with $\rho_c < \rho_c^{\star}$ would \emph{eventually} lead to
a stable stellar configuration. As a result, the head-on collision of two
neutron stars near the critical threshold can be seen as a transition in
the space of configurations from an initial stable solution over to a
critical metastable one which can either migrate to a stable solution or
collapse to a black hole~\cite{Radice:10}. As the critical limit is
approached, the survival time of the metastable object, $\tau_{\rm eq}$,
increases as $\tau_{\rm eq} = -\lambda \ln|\rho_c - \rho_c^{\star}|$,
with $\lambda \sim 10$~\cite{Jin:07a,Kellermann:08a}.

Although the free-fall velocities considered
in~\cite{Jin:07a,Kellermann:08a} were very small, the critical behaviour
continues to hold also when the stars are boosted to ultrarelativistic
velocities. Interestingly, the threshold $\rho^{\star}_{\rm c}$ becomes
now a function of the initial effective boost. Determining
$\rho^{\star}_{\rm c}$ becomes especially challenging as the Lorentz
factor is increased and the dynamics of the matter becomes extremely
violent, with very strong shocks and rarefaction waves. However, it was
possible to determine the threshold for all the range of initial boosts
considered, \ie $\vb \in [0,\, 0.95]$, $\gamma_b \in [1,\, 3.2]$, and
even to a reasonable accuracy, \eg $\rho_c^{\star} = ( 3.288023 \pm
0.000003 ) \times 10^{14}\,{\rm g/cm}^3$, for the initial boost of
$\vb=0.3$.
 
\begin{figure}
\begin{center}
  \includegraphics[width=8.0cm,angle=0]{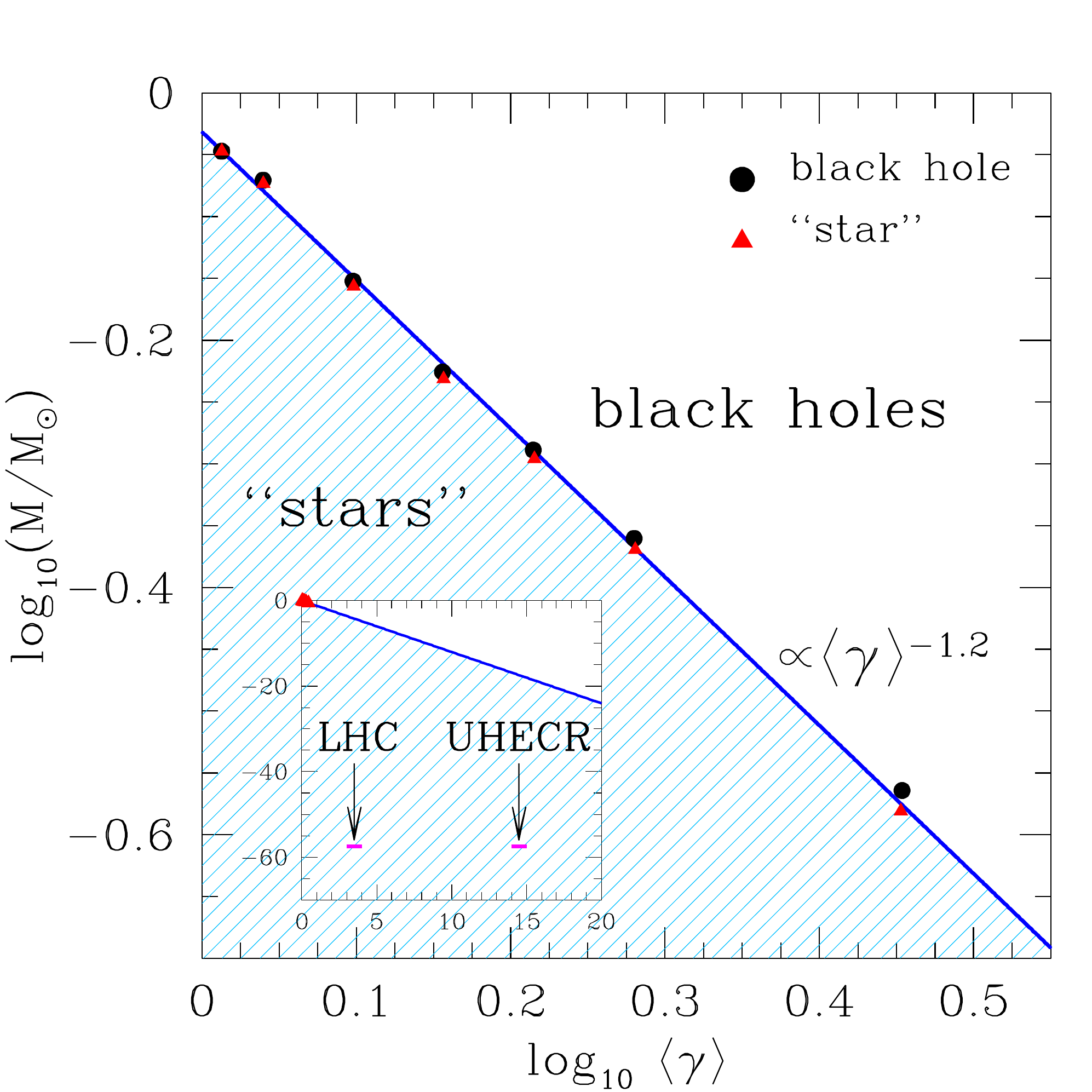}
  \newcapt{Critical line as a function of the average Lorentz factor,
    with circles indicating black holes and triangles selfgravitating
    objects. The inset shows the regimes explored at LHC and measured in
    UHECR.}
  \label{fig:fig6}
\end{center}
\end{figure}

The existence of critical behaviour near which the details of the initial
conditions become irrelevant and which is the \emph{same} at different
boosts, \ie $\lambda$ does not depend on $\gamma$ nor on $\rho_c$
(Refs.~\cite{Jin:07a,Wang2011} have shown there is ``universality'' when
varying $\gamma$ and fixing $\rho_c$), gives us a wonderful tool to
explore the conditions of black-hole formation also far away from the
masses and Lorentz factors considered in this paper. This is illustrated
in Fig.~\ref{fig:fig6}, which reports the gravitational mass of the
isolated spherical star as a function of the effective initial Lorentz
factor
\begin{equation}
\label{eq:effgamma}
\langle \gamma \rangle \equiv \frac{\int d V \, T_{\mu\nu} n^{\mu} n^{\nu}}
{(\int d V T_{\mu\nu} n^{\mu} n^{\nu})_0}\,, 
\end{equation}
where $T_{\mu\nu}$ is the stress-energy tensor, $n^{\mu}$ is the unit
normal to the spatial hyperspace with proper volume element $dV$, and the
index $0$ refers to quantities measured in the initial unboosted frame. I
should stress that the definition of the effective Lorentz
factor~(\ref{eq:effgamma}) is necessary because the stars are extended
and thus the Lorentz factor will be different in different parts of the
star. Expression~(\ref{eq:effgamma}), on the other hand, can be taken as
ratio of the energies measured in the boosted and unboosted frames, and
hence a generalisation of the Lorentz factor for a point particle (Indeed
$\langle \gamma \rangle \to 1$ for $v_b \to 1$). Of course, other
parametrizations are possible, still leading to scaling laws, but with
slightly different exponents. Filled circles indicate initial data
leading to a black hole, while triangles indicate initial data leading to
a ``star'', whereby I mean an object which is at least in part
selfgravitating (orange errorbars provide an approximate upper limit of
$\sim 8\%$ to the error in the measurements). Also indicated as a blue
solid line is the critical line separating the two regions of black hole
and star formation (the latter is shown as a shaded region). Clearly, the
numerical results provide a tight fit of the critical line with a power
law
\begin{equation}
\label{eq:Mth}
\frac{M_{\mathrm c}}{M_{\odot}} = K ~ \frac{1}{\langle \gamma \rangle^{n}} \approx 
0.92 \, \frac{1}{\langle \gamma \rangle^{1.03}}
\,.
\end{equation}

Expression~(\ref{eq:Mth}) offers itself to a number of
considerations. First, it essentially expresses the conservation of
energy. Second, in the limit of zero initial velocities, $\langle \gamma
\rangle \to 1$, one obtains that $M_{\mathrm c} \simeq 0.92 \,M_{\odot}$,
so that the corresponding total mass, $2 M_{\mathrm c}$, is only $\sim
12\%$ larger than the maximum mass of the relative spherical-star
sequence, \ie $M_{\rm max} = 1.637\,M_{\odot}$. Third, in the opposite
limit of $\langle \gamma \rangle \to \infty$, expression~(\ref{eq:Mth})
predicts that the critical mass will go zero. This is indeed what one
would expect: as the kinetic energy diverges, no room is left for
selfgravitating matter, which will all be ejected but for an
infinitesimal amount which will go into building the zero-mass critical
black hole. Fourthly, (\ref{eq:Mth}) is also in agreement with the
results in~\cite{Choptuik:2010a,East2012}, whereby one can recognize the
black-hole formation as the crossing of the critical line when moving to
larger Lorentz factors while keeping the rest-mass constant.

Finally, using~(\ref{eq:Mth}) it is possible to probe whether the kinetic
energies achieved by modern particle accelerators, such as the LHC, are
sufficient to produce micro black holes from the collision of two
ultrarelativistic particles. Using the results reported in
Ref.~\cite{Heuer_2012}, the expected energies achieved by LHC in the next
couple of years will be $4-7\,$TeV, so that a proton, whose mass is $\sim
938~\MeV \sim 8.41 \times 10^{-58}\,M_{\odot}$, can be accelerated up to
$\gamma \sim 7.5 \times 10^3$. I have therefore reported the range of
masses and Lorentz factors accessible to LHC in the inset in
Fig.~\ref{fig:fig6}, where it appears as a small magenta box. Note that
the calculations reported here do not intend to be a realistic
approximation of the dynamics of ultrarelativistic particle
collisions. However, these calculations and the presence of a critical
behaviour can be used to deduce that the ranges reachable at the LHC are
well below the critical line and thus in the region where a
partially-confined collided object is expected.

Of course, this line of arguments wildly extrapolates our results by
almost 60 orders of magnitude in mass (11 in Lorentz factor) and neglects
quantum effects and extra-dimension effects that might be important at
Planck-energy scales. Bearing in mind these caveats, our calculations
suggest that the production of micro black holes at LHC will be
unlikely. An additional confirmation that our estimates are not
unreasonable comes from considering the corresponding energy and Lorentz
factors for the observed ultra-high energy cosmic rays (UHECR), that are
observed with energies up to $\sim 10^{20}\,$eV (\ie $\gamma \sim
10^{11}$) and for which there is no evidence of black-hole formation when
interacting with the atmosphere ~\cite{Blasi2005}. Also in this case, the
relevant range of masses and Lorentz factors is shown in the inset and
falls in the region where no black holes should be produced.

As a final remark I note that the scaling relation~(\ref{eq:Mth}) can be
expressed equivalently in terms of the original stellar compactness,
$M/R$ as
\begin{equation}
\label{eq:Mth_2}
\left(\frac{M}{R}\right)_{\mathrm c} = K' ~ \frac{1}{\langle \gamma \rangle^{n'}} 
\approx 0.08 \, \frac{1}{\langle \gamma \rangle^{1.13}} \,.
\end{equation}
Since \hbox{$M_{\rm lab} \equiv \langle \gamma \rangle M$} is the mass in
the lab frame, and since $R$ is the largest dimension in that frame being
the transverse one to the motion, the ratio
\begin{equation}
\label{eq:Mth_3}
\left(\frac{M_{\rm lab}}{R}\right)_{\mathrm c} = K' 
\frac{1}{\langle \gamma \rangle^{n'-1}} 
\sim K' \frac{1}{\langle \gamma \rangle^{0.13}}\,,
\end{equation}
provides the condition for the amount of energy that, when confined in a
hoop of radius $R$, would lead to a black hole. Hence,
expression~(\ref{eq:Mth_3}) extends the spirit of the hoop conjecture to
the case in which a kinetic energy is present. Note that the limiting
value $\langle \gamma \rangle=1$ does not corresponds to a static
configuration (as in the hoop conjecture) but to a binary that is at rest
at infinity.  This explains why in this limit $({M_{\rm
    lab}}/{R})_{\mathrm c} = ({M}/{R})_{\mathrm c} \simeq 0.08$, which is
considerably smaller than the value $1/2$ predicted by the hoop
conjecture.

\subsection{Summary}

The calculations reported above demonstrate that it is possible to find a
criterion for the conditions leading to black-hole formation in the
collision of two selfgravitating fluids moving at ultrarelativistic
velocities. The Lorentz factors reached in these simulations are
considerably larger than those encountered in merging neutron-star
binaries, especially if the inspiral is along quasi-circular orbits. The
properties of the flow after the collision change with Lorentz factor,
with most of the matter being ejected in a spherical blast wave for large
boosts. Interestingly, the collided object exhibits a critical behaviour
of type I, which is found to persist also as the initial boost is
increased. This allows one to derive a simple scaling law and extrapolate
these results to the energies of elementary particles at LHC and conclude
that black-hole production is unlikely in that case.

\section{Third piece: horizons as probes of black-hole dynamics}

The third and last ``little piece'' of numerical relativity that I will
discuss is instead about calculations in vacuum spacetimes and focuses on
the merger of two black holes. This process, which represents one of the
most important source of gravitational waves, is generally accompanied by
the recoil of the final black hole as a result of anisotropic
gravitational wave emission. While this scenario has been investigated
for decades~\cite{peres:1962} and first estimates have been made using
approximated and semi-analytical methods such as a particle
approximation~\cite{1984MNRAS.211..933F}, post-Newtonian
methods~\cite{Wiseman:1992dv} and the close-limit approximation
(CLA)~\cite{Andrade:1997pc}, it is only thanks to the recent progress in
numerical relativity that accurate values for the recoil velocity have
been computed~\cite{Baker:2006nr, Gonzalez:2006md, Campanelli:2007cg,
  Herrmann:2007ac, Koppitz-etal-2007aa:shortal, Lousto:2007db,
  Pollney:2007ss:shortal, Healy:2008js, Lousto2011}.

Besides being a genuine nonlinear effect of general relativity, the
generation of a large recoil velocity during the merger of two black
holes has a direct impact in astrophysics. Depending on its size and its
variation with the mass ratio and spin, in fact, it can play an important
role in the growth of supermassive black holes via mergers of galaxies
and on the number of galaxies containing black
holes. Numerical-relativity simulations of black holes inspiralling on
quasi-circular orbits have already revealed many of the most important
features of this process showing, for instance, that asymmetries in the
mass can lead to recoil velocities $\vk \lesssim
175\ks$~\cite{Baker:2006nr, Gonzalez:2006md}, while asymmetries in the
spins can lead respectively to $\vk \lesssim 450\ks$ or $\vk \lesssim
4000\ks$ if the spins are
aligned~\cite{Herrmann:2007ac,Koppitz-etal-2007aa:shortal,
  Pollney:2007ss:shortal} or perpendicular to the orbital angular
momentum~\cite{Campanelli:2007ew, Gonzalez:2007hi,Campanelli:2007cg}
(see~\cite{Rezzolla:2008sd} for a review and \cite{Lousto2012} for the
most recent results).

At the same time, however, there are a number of aspects of the nonlinear
processes leading to the recoil that are far from being clarified even
though interesting work has been recently carried out to investigate such
aspects~\cite{Schnittman:2007ij, Mino:2008at, LeTiec:2009yg}. One of
these features, and possibly the most puzzling one, is the generic
presence of an \textit{``anti-kick''}, namely, of one (or more)
decelerations experienced by the recoiling black hole. Such anti-kicks
take place after a single apparent horizon has been found and have been
reported in essentially all of the mergers simulated so far (see Fig.~8
of Ref.~\cite{Pollney:2007ss:shortal} for some examples).

What follows discusses a phenomenological framework which provides a
novel description of the stages during which the anti-kick is generated,
and that can be used to formulate a simple and qualitative interpretation
of the physics underlying this process. I will focus on the head-on
collision of two nonspinning black holes with different mass. Although
this is the simplest scenario for a black-hole merger, it contains all
the important aspects that can be encountered in more generic
conditions\newfoot{Much of what follows is taken from the discussion
  presented in Refs.~\cite{Rezzolla:2010df, Jaramillo:2011re,
    Jaramillo:2011rf}.}.

\subsection{The basic picture}

I will start by presenting a qualitative interpretation of the antikick
by considering the simple head-on collision of two Schwarzschild black
holes with unequal masses. This qualitative picture will be made
quantitative and gauge-invariant by studying the logical equivalent of
this process in the evolution of a Robinson-Trautman spacetime, with
measurements of the recoil made at future null infinity. The insight
gained with this spacetime will be valuable to explain the anti-kick
under generic conditions and to contribute to the understanding of
nonlinear black-hole physics.

\begin{figure}
\begin{center}
  \includegraphics[width=8.0cm,angle=0]{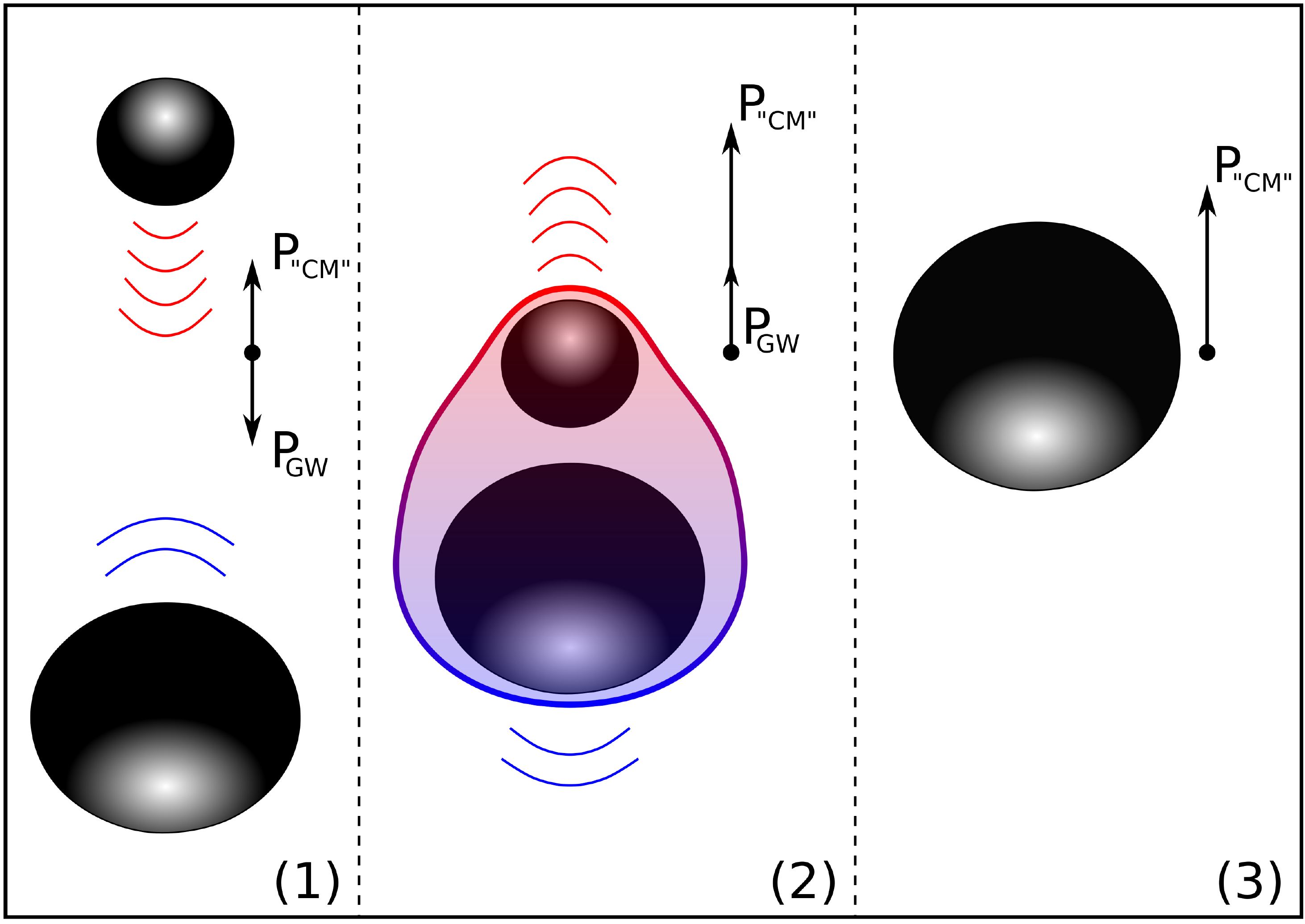}
  \newcapt{Cartoon of the generation of the anti-kick in the head-on
    collision of two unequal-mass Schwarzschild black holes. Initially the
    smaller black hole moves faster and linear momentum is radiated mostly
    downwards, thus leading to an upwards recoil of the system [stage
      (1)]. At the merger the curvature is higher in the upper hemisphere
    of the distorted black hole (\cf red-blue shading) and linear momentum is
    radiated mostly upwards leading to the anti-kick [stage (2)]. The black hole
    decelerates till a uniform curvature is restored on the horizon
    [stage (3)].}
  \label{fig:fig7}
\end{center}
\end{figure}

Figure~\ref{fig:fig7} illustrates the dynamics of the head-on collision
using a schematic cartoon where I am considering a coordinate system
centred in the total centre of mass of the system and where the smaller
black hole is initially on the positive $z$-axis, while the larger one is
on the negative axis. As the two black holes free-fall towards each
other, the smaller one will move faster and will be more efficient in
``forward-beaming'' its gravitational wave
emission~\cite{Wiseman:1992dv}. As a result, the linear momentum will be
radiated mostly downwards, thus leading to an upwards recoil of the black
hole binary [\cf stage (1) in Fig.~\ref{fig:fig7}]. At the merger, the
black-hole velocities will be the largest and so will also be the
anisotropic gravitational wave emission and the corresponding recoil of
the system. However, when a single apparent horizon is formed comprising
the two black holes, the curvature distribution on this 2-surface will be
highly \textit{anisotropic}, being higher in the upper hemisphere (\cf
red-blue shading in stage (2) of Fig.~\ref{fig:fig7}). Because the newly
formed black hole will want to radiate all of its deviations away from
the final Schwarzschild configuration, it will do so more effectively
there where the curvature is larger, thus with a stronger emission of
gravitational waves from the northern hemisphere. As a result, after the
merger the linear momentum will be emitted mostly upwards and this sudden
change in sign will lead to the anti-kick. The anisotropic gravitational
wave emission will decay exponentially as the curvature gradients are
erased and the quiescent black hole reaches its final and decelerated
recoil velocity [\cf stage (3)]\newfoot{I should remark that other
  explanations have also been suggested. One of them makes use of the
  Landau-Lifshitz pseudotensor and explains the recoil in terms of the
  cancellation of large and opposite fluxes of momentum, part of which
  are ``swallowed'' by the black hole~\cite{Lovelace:2009dg}. Another one
  is even more essential and explains the antikick is in terms of the
  spectral features of the signal at large distances, quite independently
  of the presence of a black-hole horizon~\cite{Price:2011fm}. All of
  these views serve the scope of providing an intuitive description and
  are in my view equally valid and useful.}.

Although this picture refers to a head-on collision, it is supported by
the findings in the CLA (where the direction of the ringdown kick is
approximately opposite to that of the accumulated inspiral plus plunge
kick)~\cite{LeTiec:2009yg} and it can be generalized to a situation in
which the black holes have different masses, different spins and are
merging through an inspiral. Also in a more generic case, the
newly-formed apparent horizon will have a complicated but globally
anisotropic distribution of the curvature, determining the direction
(which is in general varying in time) along which the gravitational waves
will be emitted. Hence, the geometric properties in a dynamical horizon
(of a black or white hole) determine its global dynamics. I next use the
Robinson-Trautman spacetime to validate this picture.

\subsection{A useful playground}

The Robinson-Trautman spacetime represents a class of vacuum solutions
admitting a congruence of null geodesics which are twist and
shear-free~\cite{Robinson:1962zz}, with a future stationary horizon and a
dynamical past (outer trapping) horizon~\cite{Penrose73, Tod89,
  Chow:1995va, Natorf:2008qd, Podolsky:2009an} (past apparent horizon
hereafter). A Robinson-Trautman spacetime is thus regarded as an isolated
nonspherical white hole emitting gravitational waves, where the evolution
of the apparent horizon curvature-anisotropies and the total spacetime
momentum dynamics can be related unambiguously. The metric is given
by~\cite{Macedo:2008ia}
\begin{equation}
\label{RTM}
ds^2 = -\left(K - \frac{2M_{\infty}}{r} -
\frac{2r \partial_u Q}{Q} \right) du^2 - 2dudr + \frac{r^2}{Q^2} d\Omega^2\,,
\end{equation}
where $Q=Q(u,\Omega)$, $u$ is the standard null coordinate, $r$ is the
affine parameter of the outgoing null geodesics, and $\Omega=\{
\theta, \phi \}$ are the angular coordinates on the unit sphere
$S^2$. Here $M_{\infty}$ is a constant and is related to the
asymptotic mass, while the function $K(u,\Omega)$ is the Gaussian
curvature of the surface corresponding to $r = 1$ and $u = {\rm
  constant}$, $K(u,\Omega)\equiv Q^2(1+\nabla^2_{\Omega}{\ln{Q}})$,
where $\nabla^2_{\Omega}$ is the Laplacian on $S^2$. The Einstein
equations then lead to the evolution equation
\begin{equation}
\label{RTeq}
\partial_u Q(u,\Omega)=-{Q^3} \nabla^2_{\Omega}K(u,\Omega)/({12M_{\infty}})\,.
\end{equation}
Any regular initial data $Q=Q(0,\Omega)$ will smoothly evolve according
to~(\ref{RTeq}) until it achieves a stationary configuration
corresponding to a Schwarzschild black hole at rest or moving with a
constant speed~\cite{Chrusciel:1992cj}. Equation~(\ref{RTeq}) implies the
existence of the constant of motion $\mathcal{A}\equiv
\int_{S^2}{d\Omega}/{Q^2}$, which clearly represents the area of the
surface $u={\rm const.},\,r={\rm const.}$ and can be used to normalise
$Q$ so that $\mathcal{A}=4\pi$. All the physically relevant information
is contained in the function $Q(u,\Omega)$, and this includes the
gravitational radiation, which can be extracted by relating $Q(u,\Omega)$
to the radiative part of the Riemann
tensor~\cite{deOliveira:2004bn,Aranha:2008ni}.

The past apparent horizon radius $R(u,\Omega)$ is given by the vanishing
expansion of the future ingoing null geodesics,
satisfying~\cite{Penrose73,Tod89}
\begin{equation}
\label{black holeRT}
Q^2\nabla_{\Omega}^{2} \ln{R}=K-{2M_{\infty}}/{R}\,.
\end{equation}

The mass and momentum of the black hole are computed at future null infinity
using the Bondi four-momentum~\cite{Macedo:2008ia}
\begin{equation}
\label{BondiMoment}
P^{\alpha}(u) \equiv
\frac{M_{\infty}}{4\pi}\int_{S^2}\frac{\eta^{\alpha}}{Q^3}d\Omega\,,
\end{equation}
with $\left(\eta^{\alpha}\right)= \left(1, \sin{\theta}\cos{\phi},
\sin{\theta}\sin{\phi}, \cos{\theta} \right)$. Given smooth initial data,
the spacetime will evolve to a stationary non-radiative solution which,
in axisymmetry, has the form $Q(\infty, \theta)= {\left(1 \mp v x\right)}
/ {\sqrt{1-v^2}}$, with $x\equiv \cos \theta$~\cite{Macedo:2008ia}. The
Bondi four-momentum associated to $Q(\infty, \theta)$ has components
\begin{equation}
\label{Bondiinfty}
\left(P(\infty)\right)^{\alpha}= \left({M_{\infty}}/{\sqrt{1-v^2}}\right)
\left(1,0,0,\pm v \right)\,,
\end{equation}
so that the parameter $v$ in $Q(\infty, \theta)$ can be interpreted as
the velocity of the Schwarzschild black hole in the $z$-direction.

One of the difficulties with Robinson-Trautman spacetimes is the
definition of physically meaningful initial data. Although this is meant
more as a proof-of-principle than a realistic configuration, it is
possible to adopt the prescription suggested in Ref.~\cite{Aranha:2008ni}
\begin{equation}
\label{Q0HeadOn}
Q(0,\theta)=Q_0\left[\frac{1}{\sqrt{1 - w x}}+
\frac{q}{\sqrt{1 + w x}} \right]^{-2}\,,
\end{equation}
and which was interpreted to represent the final stages (\ie after a
common apparent horizon is formed) of a head-on collision of two boosted
black holes with opposite velocities $w$ and mass ratio
$q$~\cite{Aranha:2008ni}. In practice, to reproduce the situation shown
in Fig.~\ref{fig:fig7}, it is sufficient to choose $w<0$ and take $q\in
[0,1]$; a more general class of initial data and the corresponding
phenomenology can be found in~\cite{Jaramillo:2011re,Aranha2012}. Note
that $Q_0$ is chosen so that to $\mathcal{A}=4\pi$ and that in general
the deformed black hole will not be initially at rest. As a result, given
the initial velocity $v_0\equiv P^3(0)/P^0(0)$, a boost is performed
transformation $\overline{P}^{\alpha} =
\Lambda^{\alpha}_{~\beta}(v_0)P^{\beta}$ so that $\overline{P}^{3}(0)=0$
by construction. The numerical solution of eq.~(\ref{RTeq}) with initial
data~(\ref{Q0HeadOn}) is performed using a Galerkin decomposition as
discussed in detail in~\cite{Macedo:2008ia}.

\begin{figure}
\begin{center}
  \includegraphics[width=8.0cm,angle=0]{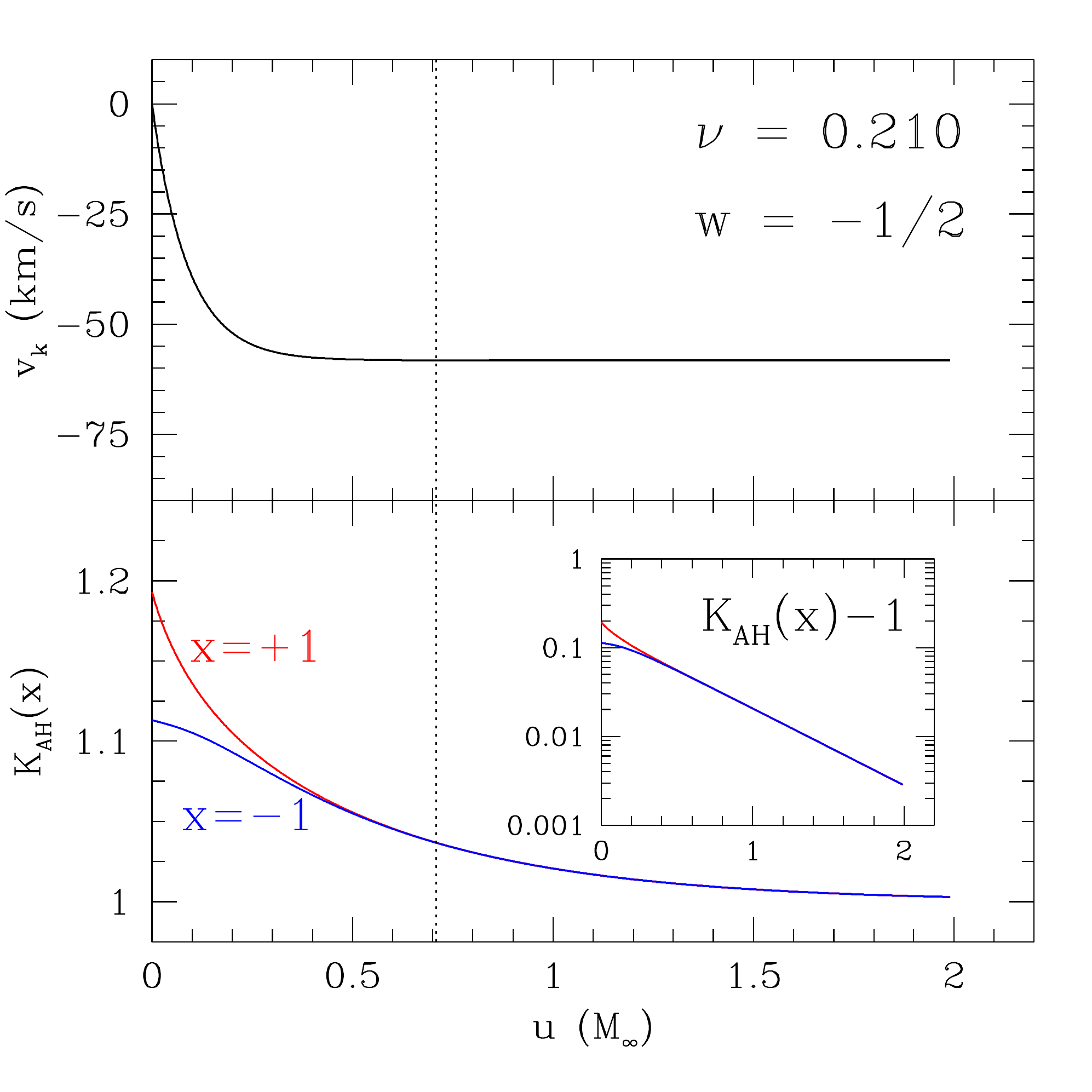} \newcapt{Typical
    evolution of a Robinson-Trautman spacetime. Shown in the lower panel is the
    evolution of the curvature $K_{_{\rm AH}}$ at the north ($x=1)$ and
    south pole ($x=-1)$. Shown in the upper panel is the evolution of the
    recoil, which stops decreasing when the curvature difference is
    erased by the emitted radiation (dotted line). Note that the
    curvature decays exponentially to that of a Schwarzschild black hole
    (inset).}  \vglue-0.5cm
  \label{fig:fig8}
\end{center}
\end{figure}

Figure~\ref{fig:fig8} reports the typical evolution of a
Robinson-Trautman spacetime with the lower panel showing the evolution of
the curvature of the past apparent horizon $K_{_{\rm AH}}\equiv
2M_{\infty}/R^3(x)$ at the north ($x = 1)$ and south pole ($x=-1)$, and
with the upper panel showing the evolution of the recoil velocity. Note
that the two local curvatures are different initially, with the one in
the upper hemisphere being larger than the one in the lower hemisphere
(\cf~Fig.~\ref{fig:fig7}). However, as the gravitational radiation is
emitted, this difference is erased. When this happens, the deceleration
stops and the black hole attains its asymptotic recoil velocity. The
inset reports the curvature difference relative to the asymptotic
Schwarzschild one, $K_{_{\rm AH}}-1$, whose exponentially decaying
behaviour is the one expected in a ringing black hole (see also Fig. 7 of
Ref.~\cite{Jaramillo:2011re}).

As mentioned before, the one shown in Fig.~\ref{fig:fig8} is a typical
evolution of a Robinson-Trautman spacetime and is not specific of the
initial data~(\ref{Q0HeadOn}). By varying the values of $w$, in fact, it
is possible to increase or decreases the final recoil and a sign change
in $w$ simply inverts the curvature at the poles so that, for instance,
initial data with $w>0$ would yield a black hole accelerating in the
positive $z$-direction. Interestingly, it is even possible to fine-tune
the parameter $w$ so that the recoil produced for a Robinson-Trautman
spacetime mimics the anti-kick produced by the quasi-circular inspiral of
nonspinning binaries. This is shown in Fig.~\ref{fig:fig1}, which reports
the recoil as a function of the symmetric mass ratio $\nu\equiv
q/(1+q)^2$, and where the dashed line refers to the anti-kick for the
inspiral of nonspinning binaries in the CLA~\cite{LeTiec:2009yg} (the
parameters chosen, \ie $w=-0.425$ and $r_{12}=2\,M$, are those minimising
the differences). Considering that the two curves are related only
logically and that the CLA one contains all the information about
inspiralling black holes, including the orbital rotation, the match is
surprisingly good.

\begin{figure}
\begin{center}
  \includegraphics[width=8.0cm,angle=0]{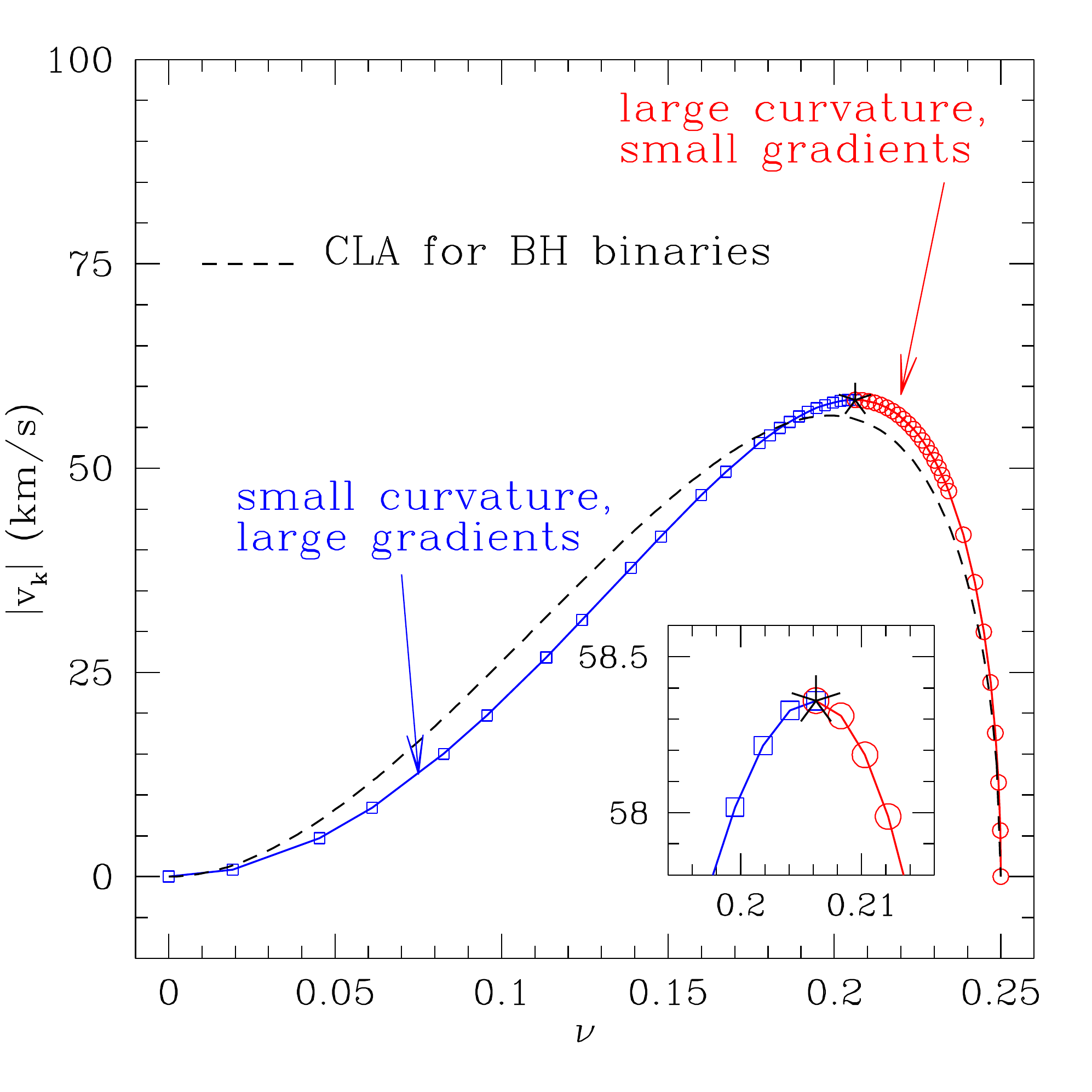}
  \newcapt{Recoil velocity shown as a function of the symmetric mass
    ratio $\nu$ when $w=-0.425$, with the dashed line refers to the
    anti-kick from the inspiral of nonspinning binaries in the
    CLA~\cite{LeTiec:2009yg}. Note that the curve can be thought as
    composed of two different branches.}
  \label{fig:fig9}
\end{center}
\end{figure}

It is also suggestive to think that the curve in Fig.~\ref{fig:fig9} is
actually composed of two different branches, one of which is
characterized by large curvature gradients across the apparent horizon
but small values of the curvature (this is the low-$\nu$ branch and is
indicated with squares), while the other is characterized by small
curvature gradients and large values of the curvature (this is the
high-$\nu$ branch and is indicated with circles). The same recoil
velocity can then be produced by two different values of $\nu$, for which
the effects of large curvature gradients and small local curvatures are
the same as those produced by small curvature gradients but large local
curvatures.

To go from this intuition to a mathematically well-defined measure one
can compute the mass multipoles of the intrinsic curvature of the initial
data using the formalism developed in~\cite{Ashtekar04a} for dynamical
horizons. Namely, it is possible to compute the mass moments as (the
mass-current are obviously zero)
\begin{equation}
\label{curv_multipoles}
M_n \equiv \oint \frac{P_{n}(\tilde{x})}{Q^2(\theta)R(\theta)}d\Omega\,,
\end{equation}
where $P_{n}(\tilde{x})$ is the Legendre polynomial in terms of the
coordinate $\tilde{x}(\theta)$ which obeys $\partial_{\theta}\tilde{x}=-
\sin\theta R(\theta)^2/(R_{_{\rm AH}}^2 Q(\theta)^2)$, with $R_{_{\rm
    AH}}\equiv \sqrt{{\cal A}_{_{\rm AH}}/(4\pi)}$ and
$\tilde{x}(0)=1$. Using these multipoles it is possible to construct an
effective-curvature parameter $K_{\rm eff}$ that represents a measure of
the global curvature properties of the initial data and from which the
recoil depends in an injective way. Because this effective-curvature
parameter has to contain the contribution from the even and odd
multipoles, the expression 
\begin{equation}
\label{Keff_AH}
K_{\rm eff} = M_2|\sum_{n=1} M_{2n+1}/3^{n-1}|\,,
\end{equation}
was found to reproduce exactly what expected (note $M_1=0$ to machine
precision).

\begin{figure}
\begin{center}
  \includegraphics[width=8.0cm,clip=true]{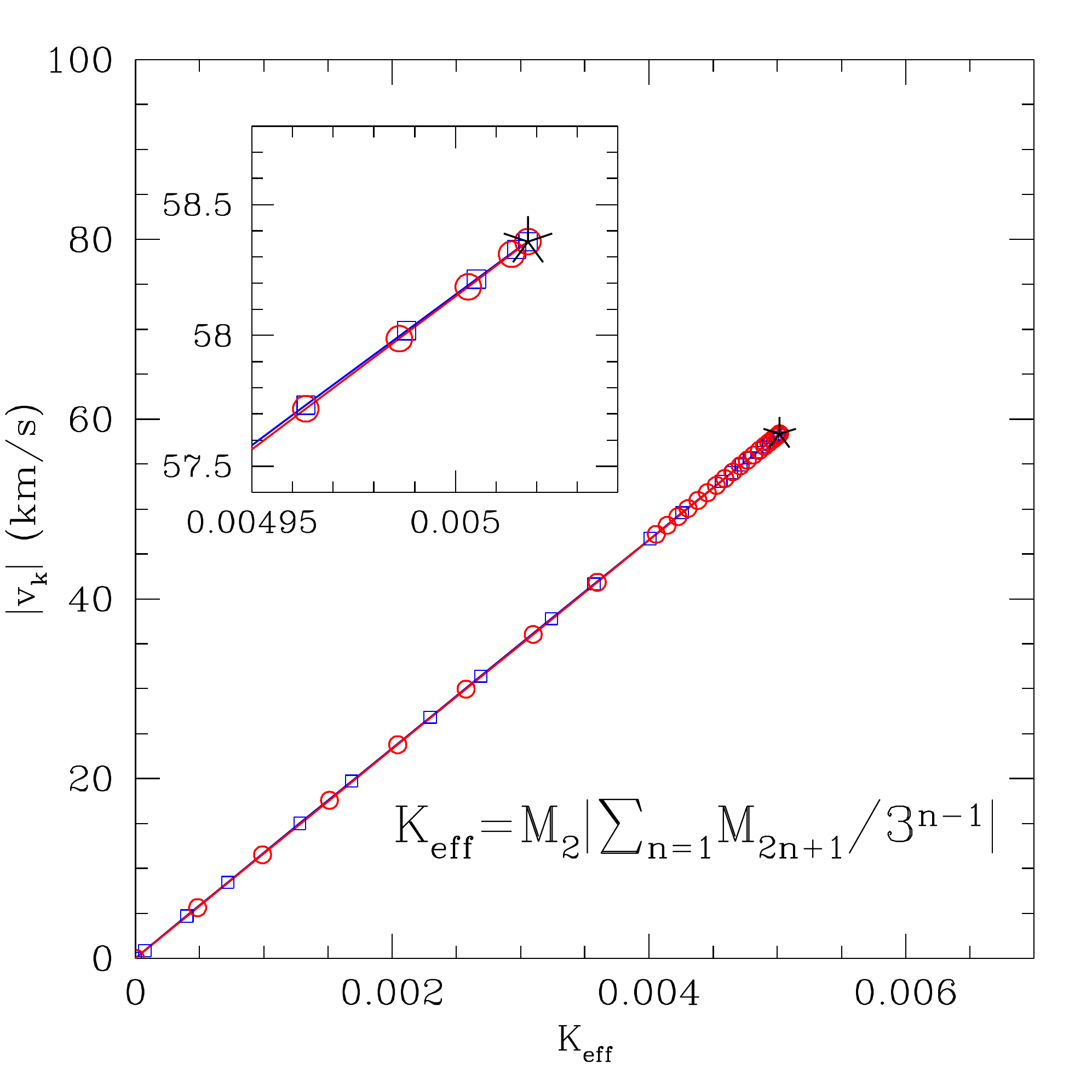}
  \newcapt{Recoil velocity shown as a function of the effective
    curvature. In contrast with Fig.~\ref{fig:fig9}, which uses the same
    symbols employed here, the relation between the curvature and the
    recoil is now injective.}  \vglue-0.5cm
  \label{fig:fig10}
\end{center}
\end{figure}

This is shown in Fig.~\ref{fig:fig10}, which reports the recoil velocity
as a function of $K_{\rm eff}$. As predicted, and in contrast with
Fig.~\ref{fig:fig9}, the relation between the curvature and the recoil is
now injective, with the maximum recoil velocity being given by the
maximum value of $K_{\rm eff}$ (see inset), and with the two branches
coinciding. The expression~(\ref{Keff_AH}) suggested above for $K_{\rm
  eff}$ is not unique and indeed a more generic one will have to include
also the mass-current multipoles to account for the spin contributions
(see discussion below). However, lacking a rigorous mathematical
guidance, the phenomenological $K_{\rm eff}$ is a reasonable, intuitive
approximation.

\subsection{A more general view}

Despite the valuable insight, the treatment summarised above and
presented in Ref.~\cite{Rezzolla:2010df} had obvious limitations. First,
the Ansatz~(\ref{Keff_AH}) for $K_{\mathrm{eff}}$, \ie $K_{\mathrm{eff}}
= f_{\mathrm{even}}\left(M_{2\ell}\right) \times
f_{\mathrm{odd}}\left(M_{2\ell+1} \right)$ is not straightforwardly
generalize to the non-axisymmetric case. Second, the functions
$f_{\mathrm{even}}$ and $f_{\mathrm{odd}}$ can be written in the simplest
possible form, \ie as a linear expansion in $M_\ell$'s, \ie
$K_{\mathrm{eff}} = \left(a_2 M_2 + a_4 M_4 +\ldots \right) \times
\left(a_3 M_3 + a_5 M_5 + \ldots \right)$, where the phenomenological
coefficients $a_\ell$'s depend on the details of the employed initial
data. Finally, the white-hole horizon analysis in Robinson-Trautman
spacetimes needs to be extended to the genuine black-hole horizon case.

While the focus in what discussed above (and presented in
Ref.~\cite{Rezzolla:2010df}) was on expressing the difference between the
{\em final} kick velocity $v_{\infty}$ and the instantaneous kick
velocity $v_{\mathrm{k}}(u)$ at an (initial) given time $u$, in terms of
the geometry of the common apparent horizon at that time $u$, it is
possible to derive a more generic view based on geometric quantities that
are evaluated at a given time during the evolution. More specifically, it
is possible to consider the variation of the Bondi linear momentum vector
in time $(dP_i^{\mathrm{B}}/du)(u)$ as the relevant geometric quantity to
monitor at null infinity $\scri^+$. This quantity can then be correlated
with a counterpart on the black-hole horizon ${\cal H}^+$, \eg a vector
$\tilde{K}^i_{\mathrm{eff}}(v)$ (function of an advanced time $v$), which
represents an extension of the effective curvature introduced in the
previous Section\newfoot{Another appealing approach that has a similar
  goal of correlating strong-fields effects with (the visualization of)
  spacetime curvature has been proposed recently by the group in
  Caltech~\cite{OweBriChe11,Nichols2011}.}.

In the case of a Robinson-Trautman spacetime, the causal relation between
the white-hole horizon ${\cal H}^-$ and null infinity $\scri^+$ made
possible to establish an explicit functional relation between
${dv_{\mathrm{k}}}/{du}$ and $K'_{\mathrm{eff}}(u)$. In the case of a
generic black-hole horizon, such a direct causal relation between the
inner horizon and $\scri^+$ is lost. However, since the corresponding
causal pasts of $\scri^+$ and ${\cal H}^-$ coincide in part, non-trivial
{\em correlations} are still possible and expected.  These correlations
can be measured by comparing geometric quantities $h_{\mathrm{inn}}(v)$
at ${\cal H}^+$ and $h_{\mathrm{out}}(u)$ at $\scri^+$, both considered
here as two timeseries\newfoot{\label{f:stretching}Note that the
  meaningful definition of timeseries cross-correlations requires the
  introduction of a (gauge-dependent) relation between advanced and
  retarded time coordinates $v$ and $u$. In an initial value problem this
  is naturally provided by the 3+1 spacetime slicing by time $t$.}. In
particular, it is reasonable to take $\tilde{K}^i_{\mathrm{eff}}(v)$ as
$h_{\mathrm{inn}}(v)$ and $(dP_i^{\mathrm{B}}/du)(u)$ as
$h_{\mathrm{out}}(u)$.

\begin{figure}
\begin{center}
\includegraphics[height=8.0cm]{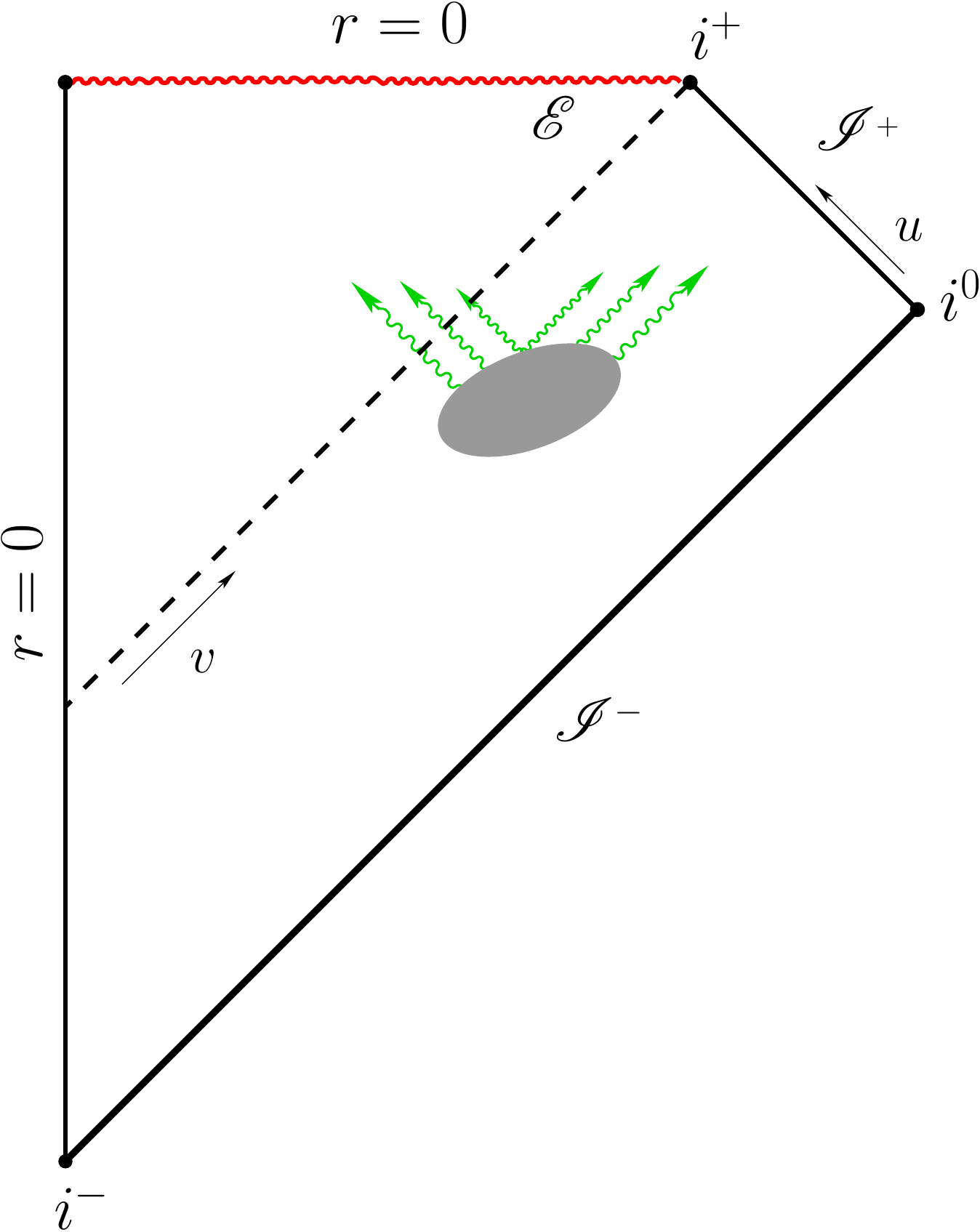}
\end{center}
\newcapt{Carter-Penrose diagram illustrating the {\em scattering}
  approach to near-horizon gravitational dynamics in a generic
  spherically symmetric collapse. The event horizon ${\cal H}^+$ and
  null infinity $\scri^+$ provide spacetime canonical screens on which
  {\em geometric quantities}, respectively accounting for horizon
  deformations and wave emission, are defined. Their cross-correlation
  encodes nontrivially information about the bulk spacetime dynamics.}
  \label{fig:fig11}
\end{figure}

This approach resembles therefore the methodology adopted in {\em
  scattering} experiments. Gravitational dynamics in a given spacetime
region affects the geometry of appropriately-chosen {\em outer} and {\em
  inner} hypersurfaces of the black-hole spacetime. These hypersurfaces
are then understood as {\em test screens} on which suitable {\em
  geometric quantities} must be constructed. The correlations between the
two encode geometric information about the dynamics in the bulk,
providing information useful for an {\em inverse-scattering} approach to
the near-horizon dynamics. In asymptotically flat black-hole spacetimes,
null infinity $\scri^+$ and the (event) black-hole horizon ${\cal H}^+$
provide natural choices for the outer and inner screens. This is
summarised in the Carter-Penrose diagram in Fig.~\ref{fig:fig11}, which
illustrates the cross-correlation approach to near-horizon gravitational
dynamics. The event horizon ${\cal H}^+$ and null infinity $\scri^+$
provide spacetime screens on which {geometric quantities}, accounting
respectively for horizon deformations and wave emission, are
measured. Their cross-correlation encodes information about the bulk
spacetime dynamics.

\begin{figure}[t]
\begin{center}
\includegraphics[height=8cm]{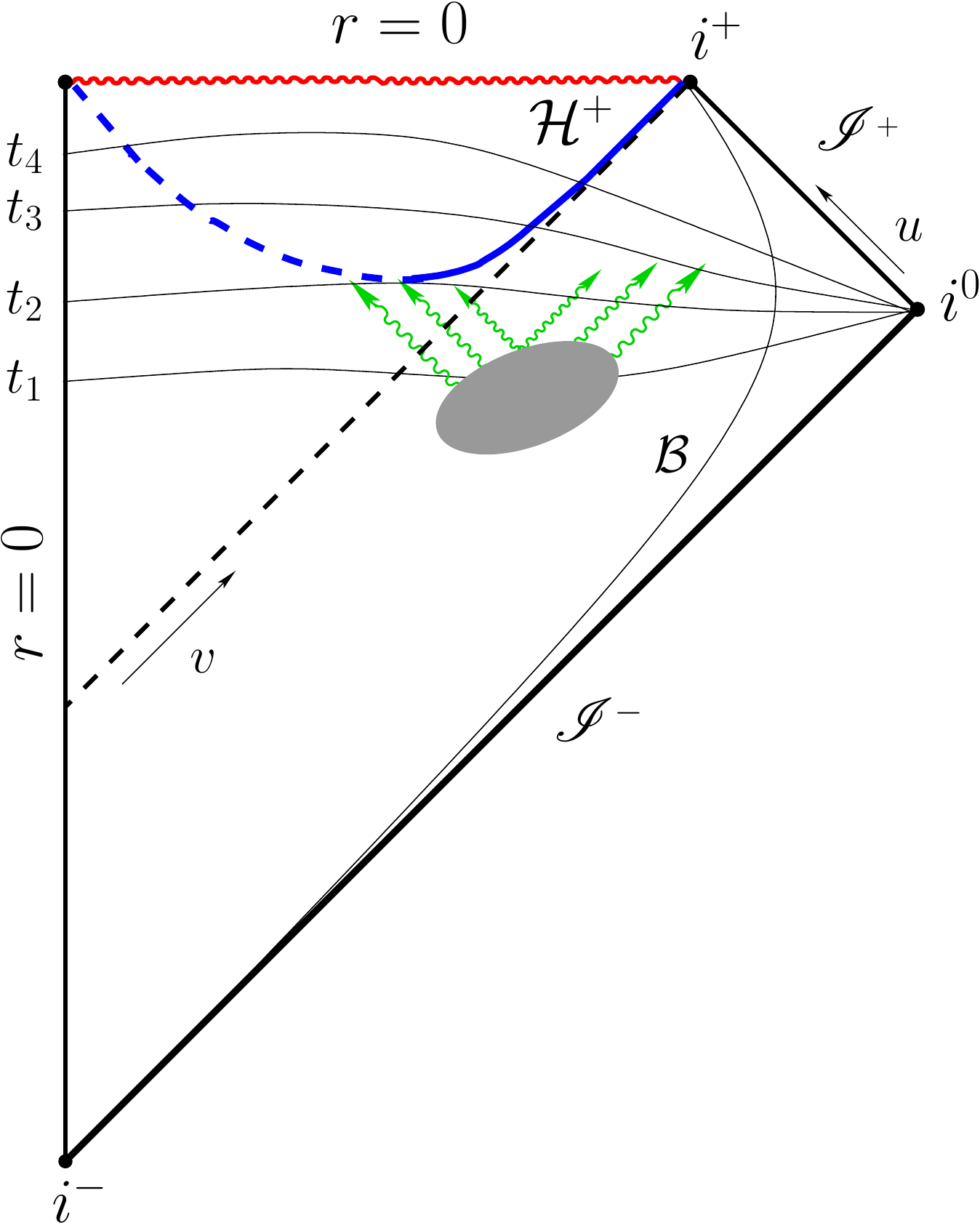}
\end{center}
\newcapt{Carter-Penrose diagram for the {\em scattering} picture in a
  Cauchy initial value approach. The dynamical horizon ${\cal H}^+$ and a
  large-distance timelike hypersurface ${\cal B}$ provide inner and outer
  screens. Note that the dynamical horizon is split in two portions:
  outer and inner (solid and dashed blue lines, respectively) and that
  the 3+1 slicing sets a common time $t$ for cross-correlations.}
\label{fig:fig12}
\end{figure}

The picture offered by Fig.~\ref{fig:fig11} can be easily adapted to the
3+1 approach commonly adopted in numerical relativity. Since neither the
black-hole event horizon nor null infinity are in general available
during the evolution\newfoot{The latter would properly require either
  characteristic or a hyperboloidal evolution approach.}, it is possible
to adopt as inner and outer screens a dynamical horizon ${\mathcal H}^+$
(future outer trapping horizon~\cite{Hayward94a, Ashtekar03a,
  Ashtekar:2004cn}) and a timelike tube ${\cal B}$ at large spatial
distances, respectively. In this case, the time function $t$ associated
with the 3+1 spacetime slicing provides a (gauge) mapping between the
{retarded} and {advanced} times $u$ and $v$, so that cross-correlations
between geometric quantities at ${\mathcal H}^+$ and ${\cal B}$ can be
calculated as standard timeseries $h_{\mathrm{inn}}(t)$ and
$h_{\mathrm{out}}(t)$. This is summarised in the Carter-Penrose diagram
in Fig.~\ref{fig:fig12}, which is the same as in Fig.~\ref{fig:fig11},
but where the 3+1 slicing sets an in-built common time $t$ for
cross-correlations between the dynamical horizon ${\cal H}^+$ (\ie the
inner screen) and a large-distance timelike hypersurface ${\cal B}$ (\ie
the outer screen).

Within this conceptual framework it is then possible to define a
phenomenological curvature vector $\tilde{K}^{\mathrm{eff}}_i(t)$ in
terms of the mass multipoles of the Ricci scalar curvature ${}^2\!R$ at
${\mathcal H}^+$ and show that this is closely correlated with a
{geometric quantities} $(dP_i^{\cal B}/dt)(t)$, representing the
variation of the Bondi linear momentum time on $\scri^+$. How to do this
in practice for a black-hole spacetime requires much more space that I
can take in this contribution and therefore refer the interested reader
to Refs.~\cite{Jaramillo:2011re,Jaramillo:2011rf}, where this is
discussed in great detail.

\subsection{Summary}

The discussion reported above demonstrates that qualitative aspects of
the post-merger recoil dynamics at infinity can be understood in terms of
the evolution of the geometry of the common horizon of the resulting
black hole. Moreover, suitably-built quantities defined on inner and
outer worldtubes (represented either by dynamical horizons or by timelike
boundaries) can act as test screens responding to the spacetime geometry
in the bulk, thus opening the way to a cross-correlation approach to
probe the dynamics of spacetime. This picture was shown to hold both for
a simple Robinson-Trautman spacetime, but also for more generic binary
black-hole spacetimes. In this latter case, this is possible through the
construction of a phenomenological vector $\tilde{K}^{\mathrm{eff}}_i(t)$
from the Ricci curvature scalar ${}^2\!R$ on the dynamical horizon
sections, which then captures the global properties of the flux of Bondi
linear momentum $(dP_i^{\mathrm{B}}/dt)(t)$ at infinity, namely the
acceleration of the BH. 

A geometric framework looking at the horizon's properties offers a number
of connections with the literature developing around the use of horizons
to study the dynamics of black holes, as well as with the interpretations
of such dynamics in terms of a viscous-hydrodynamics analogy. Much of the
machinery developed using dynamical trapping horizons as inner screens
can be extended also when a common horizon is not formed (as in the
calculations reported in Ref.~\cite{Sperhake:2010uv}). While in such
cases the identification of an appropriate hypersurface for the inner
screen can be more difficult, once this is found its geometrical
properties can be used along the lines of the cross-correlation approach
discussed here for dynamical horizons.

\section{Conclusions}

The ``three little piece'' for numerical computer and relativity
presented in the sections above ought to provide a reasonable idea of the
``Renaissance'' that numerical relativity is now experiencing. More
importantly, they should be able to convey the enormous potential that
numerical-relativity simulations have in revealing aspects of the theory
that cannot be handled analytically, or in exploring nonlinear regimes
that cannot be investigated through perturbative approaches. As remarked
repeatedly, the examples brought represent only a personal (and biased)
selection of the intense work carried out recently and surely are not
exhaustive in terms of the physical scenarios that can be explored. Much
more can be said about this and surely it will not have to wait for the
bicentenary of Einstein's stay in Prague.

\section*{Acknowledgements}
The work discussed here has been carried out in collaboration with
M. A. Aloy, L. Baiotti, B. Giacomazzo, J. Granot, J. L. Jaramillo,
C. Kouveliotou, R. P. Macedo K. Takami, whom I am indebted with. My
thanks go also to the numerical-relativity group of the AEI for providing
such a stimulating and productive environment. Support comes through the
DFG grant SFB/Trans-regio 7 and ``CompStar'', a Research Networking
Programme of the ESF. The calculations have been performed on the
clusters at the AEI.

\section*{References}
\bibliography{Prague_100years}

\end{document}